\documentclass{aa}  

\usepackage{graphicx}
\usepackage{txfonts}
\begin{document}

   \title{Probing the Dragonfish star-forming complex: the ionizing population of the young massive cluster Mercer 30\thanks{Based on observations collected at the European Organisation for Astronomical Research in the Southern Hemisphere, Chile, under programs IDs 179.B-2002, 081.D-0471, 083.D-0765, 087.D-0957, \& 089.D-0989.}}
    \titlerunning{Probing the Dragonfish: the ionizing population of Mercer 30}

    \author{D. de la Fuente\inst{1}
          \and
          F. Najarro\inst{1}
          \and
          J. Borissova\inst{2,3}
          \and
          S. Ram\'irez Alegr\'ia\inst{2,3}
          \and
          M. M. Hanson\inst{4}
          \and
          C. Trombley\inst{5}
          \and
          D. F. Figer\inst{5}
          \and
          B. Davies\inst{6}
          \and
          M. Garcia\inst{1}
          \and
          R. Kurtev\inst{2,3}
          \and
          M. A. Urbaneja\inst{7}
          \and
          L. C. Smith \inst{8}
          \and
          P. W. Lucas \inst{8}
          \and
          A. Herrero \inst{9,10}
          }

   \institute{Centro de Astrobiolog\'ia (CSIC/INTA), ctra. de Ajalvir km. 4, 28850 Torrej\'on de Ardoz, Madrid, Spain\\
              \email{delafuente@cab.inta-csic.es}
         \and
             Departamento de F\'isica y Astronom\'ia, Facultad de Ciencias, Universidad de Valpara\'iso, Av. Gran Bretaña 1111, Playa Ancha, Casilla 5030, Valpara\'iso, Chile
         \and
             Millennium Institute of Astrophysics (MAS), Santiago, Chile
         \and
              Department of Physics, University of Cincinnati, Cincinnati, OH 45221-0011, USA
         \and
             Center for Detectors, Rochester Institute of Technology, 74 Lomb Memorial Drive, Rochester, NY 14623, USA
         \and
             Astrophysics Research Institute, Liverpool John Moores University, 146 Brownlow Hill, Liverpool L3 5RF, UK
         \and
             Institute for Astro- and Particle Physics, University of Innsbruck, Technikerstr. 25/8, A-6020 Innsbruck, Austria
         \and
             Centre for Astrophysics Research, Science and Technology Research Institute, University of Hertfordshire, Hatfield AL10 9AB, UK
         \and
             Instituto de Astrofísica de Canarias, 38200 La Laguna, Tenerife, Spain
         \and
             Departamento de Astrofísica, Universidad de La Laguna, 38205 La Laguna, Tenerife, Spain
             }

   \date{Received December 18, 2015; accepted February 7, 2016}


\abstract{It has recently been claimed that the nebula, Dragonfish, is powered by a superluminous but elusive OB association. However, systematic searches in near-infrared photometric surveys have found many other cluster candidates in this region of the sky. Among these, the first confirmed young massive cluster was Mercer 30, where Wolf-Rayet stars were found.We perform a new characterization of Mercer 30 with unprecedented accuracy, combining NICMOS/HST and VVV photometric data with multi-epoch ISAAC/VLT H- and K-band spectra. Stellar parameters for most of spectroscopically observed cluster members are found through precise non-LTE atmosphere modeling with the \texttt{CMFGEN} code. Our spectrophotometric study for this cluster yields a new, revised  distance of $d = (12.4 \pm 1.7)$ kpc and a total of $Q^H_\mathrm{Mc30} \approx 6.70 \times 10^{50} \mathrm{s}^{-1}$ Lyman ionizing photons.
A cluster age of $(4.0 \pm 0.8)$ Myr is found through isochrone fitting, and a total mass of $(1.6 \pm 0.6) \times 10^4 M_\odot$ is estimated, thanks to our extensive knowledge of the post-main-sequence population. As a consequence, membership of Mercer 30 to the Dragonfish star-forming complex is confirmed, allowing us to use this cluster as a probe for the whole complex, which turns out to be extremely large ($\sim$ 400 pc across) and located at the outer edge of the Sagittarius-Carina spiral arm ($\sim$ 11 kpc from the Galactic center). The Dragonfish complex hosts 19 young clusters or cluster candidates (including Mercer 30 and a new candidate presented in this work) and an estimated minimum of nine field Wolf-Rayet stars. All these contributions account for, at least 73\% of the ionization of the Dragonfish nebula  and leaves little or no room for the alleged superluminous OB association; alternative explanations are discussed.}


   \maketitle
%

 \section{Introduction}

Young Massive Clusters (YMCs) with ages $\lesssim 10$ Myr play a fundamental role in the physics and evolution of spiral galaxies. These objects host large populations of hot massive stars that have a significant impact on the surrounding interstellar medium (ISM), by means of strong winds, photoionization, and supernovae. Initially, these feedback processes cause the rapid expulsion of intracluster gas \citep{hills80, goodwin-bastian06, weidner+07} on timescales of $\sim$ 1 - 3 Myr \citep{allen+07, walch+12, hollyhead+15}, which can lead to violent relaxation or even complete dissolution of the cluster \citep[the so-called infant mortality;][]{lada-lada03, bastian-goodwin06}.
After the natal cloud is removed, YMCs continue excavating their environment, sculpting distinctive structures such as bubbles, shells, pillars, or ionization fronts \citep{dale+12a, dale+12b, rogers-pittard13}. The mechanical energy injected into the surrounding neutral ISM may trigger the formation of new stars \citep{elmegreen-lada77, elmegreen98} or can have just the opposite effect, inhibiting star formation owing to the dispersal of molecular clouds \citep{williams-mckee97, walch+12}.

Because of the hierarchical nature of star formation \citep{elmegreen+06}, YMCs, and their ionized surroundings are commonly arranged in giant complexes of recent star formation, as observed in spiral arms of external galaxies \citep{zhang+01, larsen+04, bastian+05b, elmegreen+06, elmegreen+14}. Such regions are ideal laboratories for investigating extensively the influence of hot massive stars in their environment, especially with regard to photoionization and the triggering of star formation. In principle, this topic can be adressed in nearby face-on spiral galaxies, where YMCs are easily detected owing to their extremely high luminosities and their distinctive colors \citep[see e.g.,][]{whitmore+99,bastian+05a,scheepmaker+07,chandar+10}.
However, serious problems arise from the limited spatial resolution of extragalactic surveys, which requires using indirect methods to find out the fundamental properties of YMCs. For example, mass measurements are quite uncertain since these are based on mass-to-light ratios that are strongly dependent on ages \citep{larsen08}, and age determination methods of extragalactic YMCs are unreliable unless specific spectral types of cluster members are known \citep{hollyhead+15}. Also, the difficulties of measuring cluster sizes and dynamics prevent us from making the distinction between bound YMCs, supervirial OB associations, and extremely luminous stars \citep{silvavilla-larsen11,bastian+12}.

In this context, detailed studies of resolved Galactic YMCs within massive star-forming complexes become crucial. Unfortunately, our unfavorable location in the Milky Way limits our knowledge of such massive complexes to a small fraction of the expected total population \citep{hanson-popescu07, ivanov12}, even at infrared wavelengths, where the extinction is relatively low. Moreover, even detected cluster candidates can neither be confirmed nor characterized with imaging alone, owing to superimposed stellar populations along the line of sight: a nearby group of cool low-mass stars might show similar magnitudes and colors than a reddened massive OB association. In some cases \citep{baumgardt+98, perren+12}, overdensities that resembled clusters turn out to be chance alignments of stars at different distances. Therefore, very careful characterizations are required to confirm stellar clusters and/or associations as part of the same massive complex.

    \begin{figure}
      \centering
        \includegraphics[width=\hsize, bb=15 25 595 450, clip]{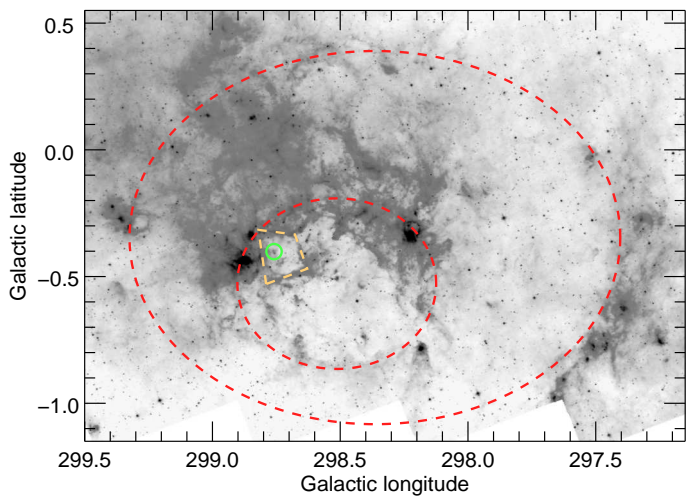}
        \caption{Image of the Dragonfish nebula in the [8.0] band of the GLIMPSE survey, indicating the cavities found by \citet{murray-rahman10, rahman-murray10} (red ellipses), as well as the box-shaped bubble (orange line) around Mercer 30 (green circle).
              }
      \label{fig:cavities}
    \end{figure}

This paper aims to investigate the ionizing sources of a giant star-forming complex in the Milky Way, focusing on a YMC whose hot massive stellar content will be analyzed in depth. The \object{G298.4-0.4} complex (the \object{Dragonfish nebula}), was first detected by \citet{russeil97} as a group of \ion{H}{ii} regions, all of them being located at a distance of $\sim 10$~kpc except one (RCW 64) that was a foreground object. In subsequent years, several cluster candidates were identified in the area by \citet{dutra+03} and \citet{mercer+05}.
Follow-up spectrocopic observations of the cluster candidate \object{Mercer 30}  (a.k.a. \object{GLIMPSE 30}; $l=298.755$, $b=-0.408$) by \citet{kurtev+07} yielded four massive stars of spectral types WR and Of, thereby confirming the YMC nature of this object. Further research  has found many additional cluster candidates \citep{borissova+11, morales+13, solin+14, barba+15}, and one of them (\object{VVV CL011}) was confirmed as hosting blue massive stars by \citet{chene+13}.

The Dragonfish region was revisited in a systematic search for the most luminous star-forming complexes in the Milky Way by \citet{murray-rahman10} and \citet{rahman-murray10}. These papers investigate microwave and infrared data from space surveys (specifically WMAP, GLIMPSE and MSX), finding many giant bubbles of ionized ISM that were interpreted as being inflated by YMCs. In relation to the Dragonfish nebula, \citet{murray-rahman10} indicate the presence of two \ion{H}{ii} cavities in the $8 \mu m$ GLIMPSE image, one enclosed within the other (see Fig. \ref{fig:cavities}, where we also show a smaller box-shaped bubble around Mercer 30 that was previously unnoticed).
The wider one, whose dimensions are coincident with the WMAP source, is mainly demarcated by a bright rim at $l \sim 297.5$ and a patch at $(l,b) \approx (299.35,-0.3)$ that we identify as the foreground object \object{RCW 64}. Close to the border of the inner elliptic cavity, five \ion{H}{ii} regions were found by \citet{murray-rahman10} with an estimated total flux of 110 Jy. On the other hand, their free-free flux estimate of the entire emitting region (i.e., all the clouds shown in Fig. \ref{fig:cavities}, including the extended emission close to the edges of the large ellipse) is 312 Jy. These authors claim that an extremely massive cluster should be responsible for the remaining 202 Jy, but they disregard the contribution of \ion{H}{ii} regions outside the inner cavity for such calculation, as well as the presence of Mercer 30 and other cluster candidates.
Following this claim, \citet{rahman+11a} find a shallow, spatially extended ($10' \times 11'$) overdensity in the 2MASS point-source catalog \citep{2mass}, considering only those colors that could correspond to OB stars under extinctions $A_K = 1$. As explained above, these photometric conditions may also correspond to less distant late-type stars; in fact, \citet{rahman+11a} consider K-type giants as contaminating stars. The overdensity, which was named Dragonfish Association, was allegedly confirmed as an OB association by \citet{rahman+11b} through low-resolution ($R \sim 1000$) H- and K-band spectroscopy.

In this paper, we  carry out an extensive spectroscopic analysis of massive stars in Mercer 30. We intend to use this YMC as a measuring probe for the ionizing stellar population in the Dragonfish complex. This procedure will allow us to reassess the source of the Dragonfish nebula's ionization, as well as the nature of the so-called Dragonfish Association.

\section{Observations and data reduction}

  \subsection{Photometry}
  \label{sec:photometry}

The photometric data we present in this paper comes from two complementary sources: \textit{Hubble} Space Telescope (HST) imaging and the VISTA Variables in the V\'ia L\'actea (VVV) public survey. The VVV images have an ample spatial coverage that includes the outskirts of the cluster, however their resolution is not sufficient to resolve the central region. Conversely, the HST images have a limited field of view but are able to resolve the most crowded regions of Mercer 30. Below we adress these two photometric datasets separately.

Mercer 30 was observed on 2008 July 18 with the Near Infrared Camera and Multi-Object Spectrometer (NICMOS) onboard the HST, as part of the observing program \#11545 (PI: Davies). Images centered at Mercer 30 were taken with the NIC3 camera, whose pixel scale is $0.2''$, using the F160W and F222M filters to obtain broad-band photometry. In addition, the same field was observed with the F187N and F190N narrow-band filters, which are coincident with the Paschen-$\alpha$ line and the adjacent continuum. The observing strategy includes a 6-point dither pattern that allows a better sampling of the point-spread function.
Data reduction was carried out following the NICMOS Data Handbook v7.0 and using the \texttt{calnica} software, which is specifically built for NICMOS imaging. The reduction process involves bias and dark-current subtraction, flat-field correction, pixel resampling onto a finer grid, and combination of frames in a single image. Photometry was extracted using an IDL-adapted version of \texttt{DAOPHOT} \citep{stetson87}. Due to the crowded nature of the field, a small aperture of $r < 0.4''$ was preferred, along with an annulus of $1.3'' < r < 2.0''$ for background subtraction.

To perform the VVV photometry of the cluster we used the VVV-SkZ pipeline \citep{mauro+13}, an automated software based on \texttt{ALLFRAME} \citep{stetson94}, and optimized for VISTA PSF photometry. We measured the $J$ and $H$ photometry over the stacked images, observed on 27 March, 2010, each with an exposure time of 40 seconds and downloaded from the VISTA Science Archive (VSA) website. The $K_S$ photometry was calculated directly from the stacked images observed between 14 March, 2010 and 13 May, 2013 (68 images). We calibrated the VVV instrumental photometry using 2MASS stellar sources in the image. Photometric errors are lower than 0.2 mag for $K_S<19$ mag, and for saturated bright stars ($K_S<9.5$ mag) we used 2MASS photometry.

\begin{table*}
  \caption{Summary of spectroscopic observations of Mercer 30}             
  \label{tab:obs}      
  \centering          
  \begin{tabular}{c c c c c c}       
  \hline\hline 
Program ID & P.I. & Start date\tablefootmark{a} & End date\tablefootmark{a} &  Slit widths & Central wavelengths $[\mu\mathrm{m}]$ \\
  \hline
081.D-0471 & Borissova    & 2008-06-13 & 2008-06-15 & $0.6''$           & 1.705, 2.15 \\
083.D-0765 & Puga         & 2009-04-14 & 2009-04-16 & $0.3''$, $0.8''$  & 1.71, 2.09, 2.21 \\
087.D-0957 & de la Fuente & 2011-04-18 & 2011-05-14 & $0.8''$           & 1.71, 2.09, 2.21 \\
089.D-0989 & de la Fuente & 2012-03-13 & 2012-03-13 & $0.8''$           & 2.09, 2.21 \\
  \hline                  
  \end{tabular}
  \tablefoot{
  \tablefoottext{a}{Dates refer only to the Mercer 30 observations presented in this paper, and not to the whole programs, which also included other clusters.}
  }
\end{table*}

  \subsection{Spectroscopy}
    \label{sec:spectroscopy}

We selected for spectroscopy those targets that have Paschen-$\alpha$ emission (i.e. F187N > F190N), as well as other bright stars in the cluster field. Emission in Paschen-$\alpha$ has  proven very effective to pinpoint hot massive stars in clusters \citep{davies+12a, davies+12b, delafuente+13, delafuente+15}.

Spectroscopic observations were carried out at the 8.2-m Unit 1 telescope of the ESO Very Large Telescope (VLT), located on Cerro Paranal in Atacama, Chile, with the Infrared Spectrometer Array Camera \citep[ISAAC;][]{moorwood+98}, under several campaigns that are listed in Table \ref{tab:obs}. Spectra were obtained at H and K bands with different central wavelengths (see also Table \ref{tab:obs}).
We used three slits whose widths are 0.8, 0.6, and 0.3 arcseconds, providing spectral resolving powers of $R\sim$ 4000, 5000, and 10000, respectively. Within each slit position, we aligned two or more stars of similar brightness. Owing to stellar crowding, additional stars were unintentionally observed, yielding several bonus spectra at various distances to the cluster center. To provide background subtraction, targets were observed in two offset positions, A and B, and additional small dithering around A and B was performed.

Data reduction was carried out independently by two research groups that used different techniques. Below we explain both reduction procedures separately.

 The reduction process has already been explained by \citet{hanson+10} for the 2008 observations (program ID: 081.D-0471), therefore we refer to that paper for details, and we  provide here only a brief summary. Following the ISAAC Data Reduction Guide (Version 1.5) and using IRAF together with the ESO \texttt{eclipse} package \citep{devillard01}, the images were first corrected from ghost images, flat field, and warping. Spectra were extracted and wavelength-calibrated using Xenon and Argon arc-lamps. To remove the atmosphere features, telluric spectra were obtained by modeling and removing the hydrogen lines from the spectrum of the nearby A0V star HD 106797.

For the remaining observing programs listed in Table \ref{tab:obs}, reduction was carried out through a custom-built IDL pipeline. The first step of the reduction process consists of correcting the warp of the two-dimensional frames. While distortion along the spatial axis is estimated using the STARTRACE frames that are part of the ISAAC calibration plan, warping along the spectral axis is calculated by fitting the OH emission lines that are imprinted in the science frames. These OH lines are also used for wavelength calibration, taking the wavelengths in vacuum from \citet{rousselot+00}. The wavelength residuals have a root mean square of $\sim 0.5$ \AA, which roughly corresponds to one tenth of the resolution element. We apply the resulting rectification matrix to both flat field and science frames, and then the latter are divided by the normalized average flat field. The next step consists of correcting sky background, taking advantage of the aforementioned A and B offset positions.
We subtract each AB or BA pair, yielding sky-cleaned positive and negative spectra. All these spectra are then extracted and combined to obtain a one-dimensional spectrum for each object. This includes spectra of telluric standards (late-B dwarfs) that were taken immediately before or after each programmed observing block and at a similar airmass, as part of the calibration plan provided by ESO. As the only H- and K-band intrinsic features of these standard stars are hydrogen absorption lines (specifically, the Brackett-series lines Br$\gamma$, Br10, Br11), pure telluric spectra are obtained simply by removing such features through Voight-profile fitting. After dividing the spectra of all our targets by the corresponding telluric spectra, normalization is carried out by continuum-fitting with 3rd to 5th degree polynomials.

The wavelength axis of each spectrum has been shifted to subtract the observatory motion in the radial direction, using the \texttt{rvcorrect} task of IRAF. As a result, all the final spectra are set in the Local Standard of Rest (LSR) reference frame.

    \begin{figure*}
      \centering
        \includegraphics[height=8cm]{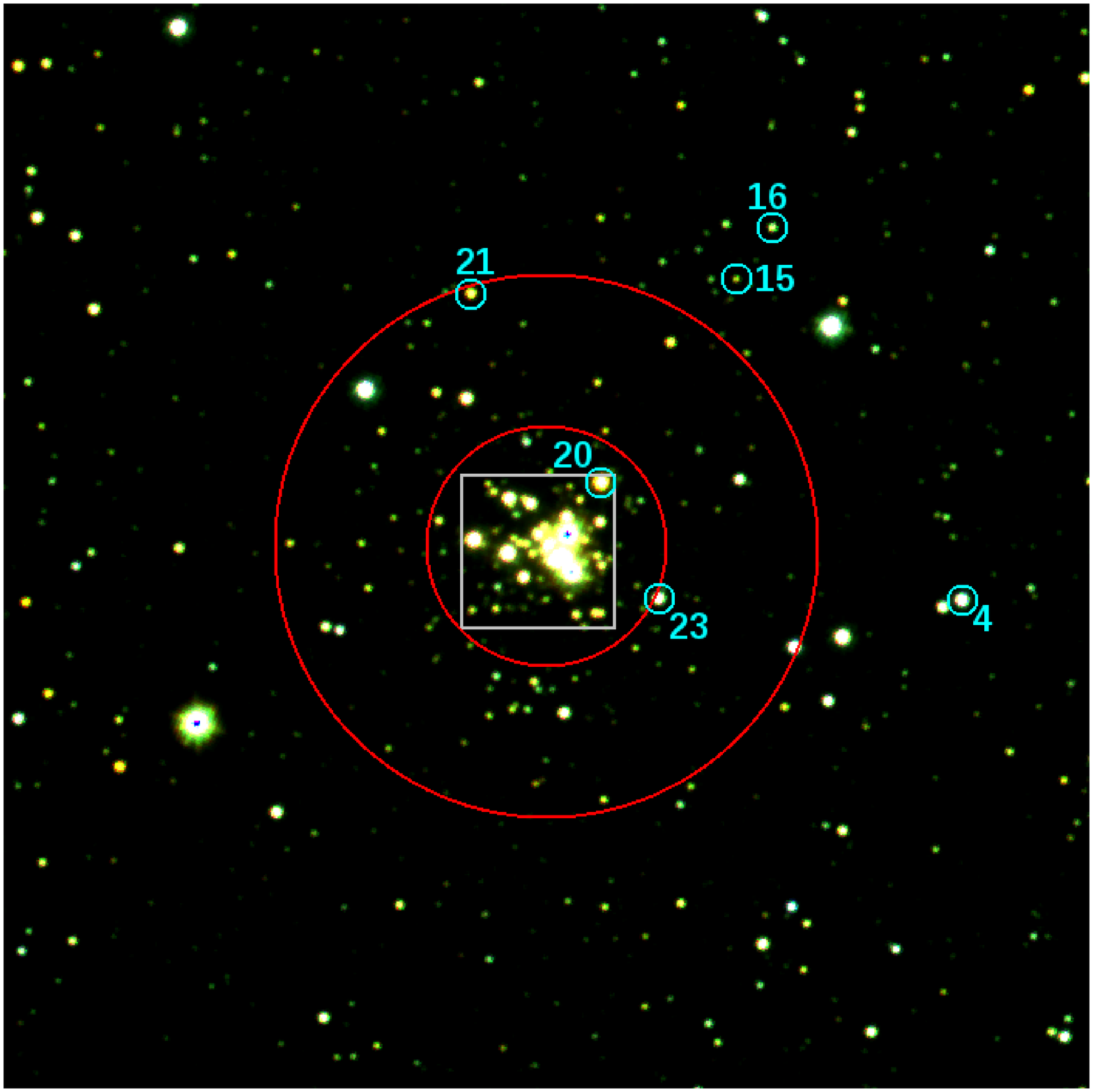}
        \qquad
        \includegraphics[height=8cm, bb=30 12 635 595, clip]{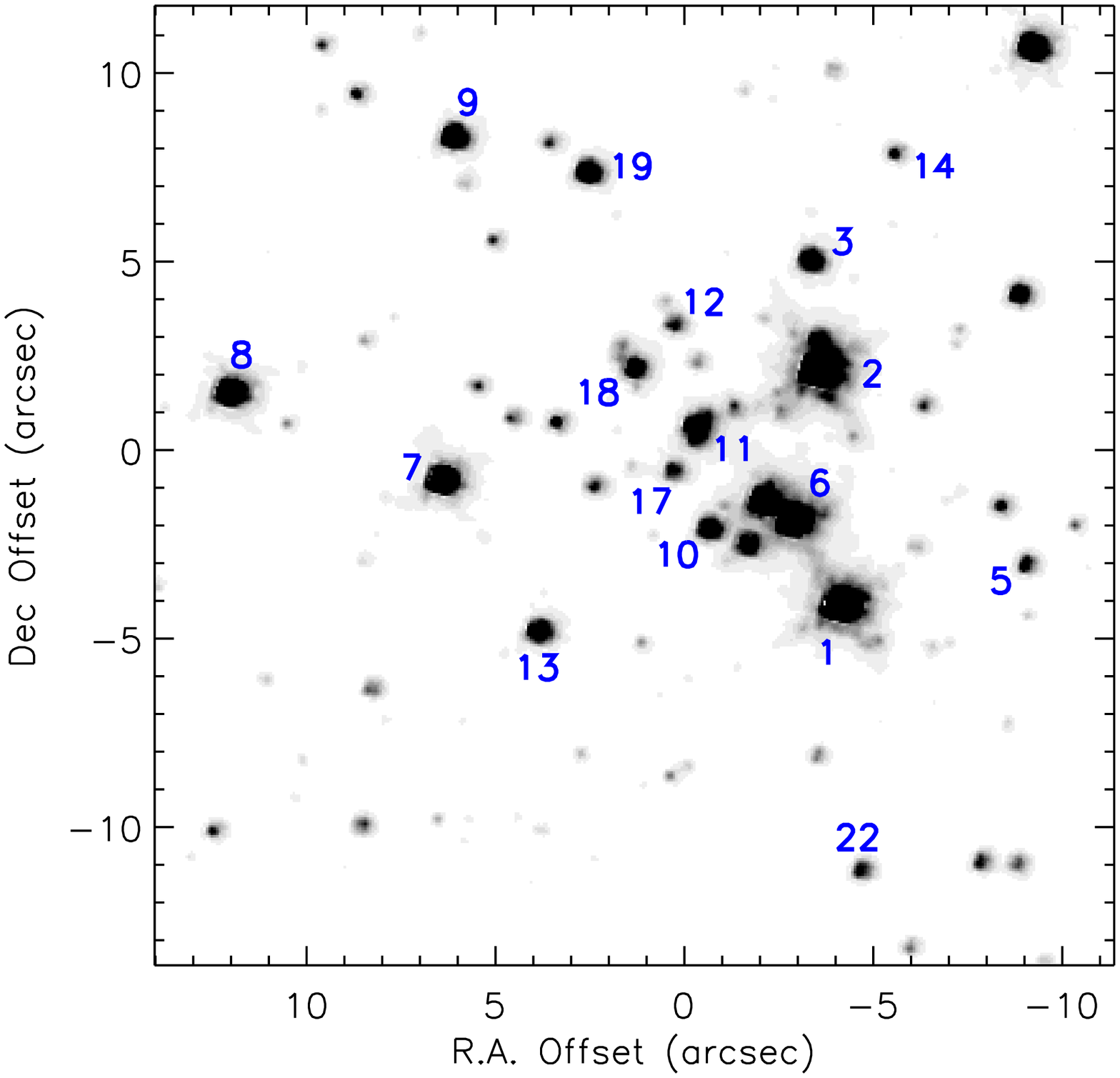}
        \caption{$3' \times 3'$ RGB image (R $=K_S$, G $=H$, B $=J$) of Mercer 30 from the VVV survey (left) and NICMOS/HST close-up view of the central region in the F222M band (right). The spatial coverage of the latter is shown as a gray square on the VVV image. The origin of the coordinate plane in the HST chart corresponds to the center of the VVV image, which is situated at R.A. $ = 12^{h} 14^{m} 32.15^{s}$, Dec. $= -62\degr 58' 50.1''$. North is up and east is left. All the spectroscopically observed stars are labeled as identified in Table \ref{tab:stars}. Confirmed cluster members are shown only in the HST chart and the remaining objects are labeled in the VVV image. Red circles enclose the regions $r < 20''$ and $20'' < r < 45''$, where the HST and the VVV photometry, respectively, have been used for cluster characterization (see section \ref{sec:cmd})}
      \label{fig:finders}
    \end{figure*}

\section{Analysis and characterization of Mercer 30}

Although a first study was carried out by \citet{kurtev+07}, our extensive spectroscopy and new photometric data will considerably improve  our knowledge of Mercer 30, and especially of its massive population. As we  show in subsequent sections, our improvements are mainly due to higher spatial resolution of images for the central regions of the cluster as well as the large number of spectroscopically observed stars.

    \begin{figure*}
      \centering
        \includegraphics[width=0.75\hsize]{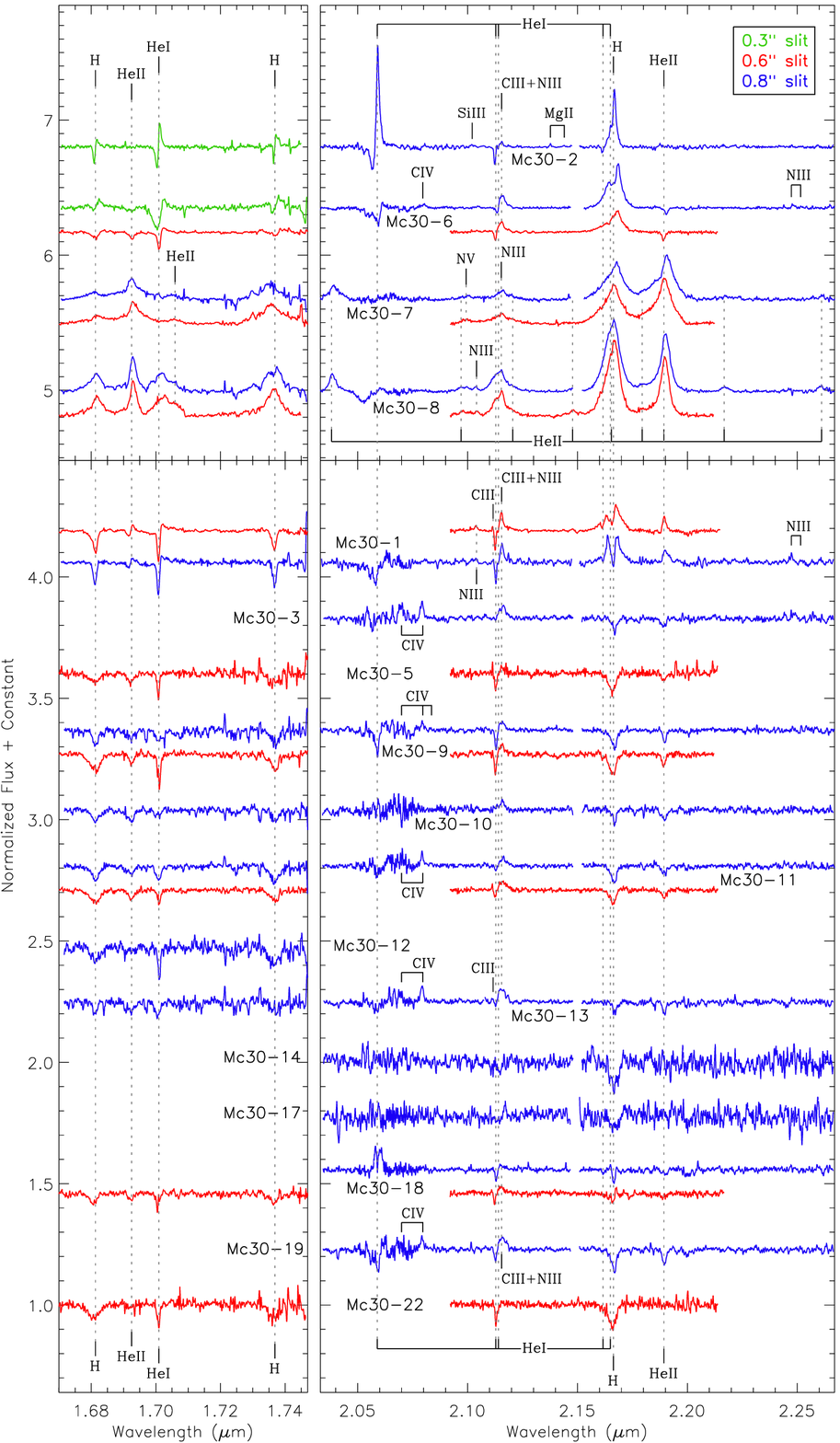}
        \caption{H- and K-band spectra of hot stars in Mercer 30. Note the different scale of the upper panels to properly display  strong emissions. Wavelengths of identification marks have been shifted to match the radial velocity of the cluster.}
      \label{fig:hotspectra}
    \end{figure*}

  \subsection{Spectral classification}
  
\begin{figure}
      \centering
        \includegraphics[width=\hsize]{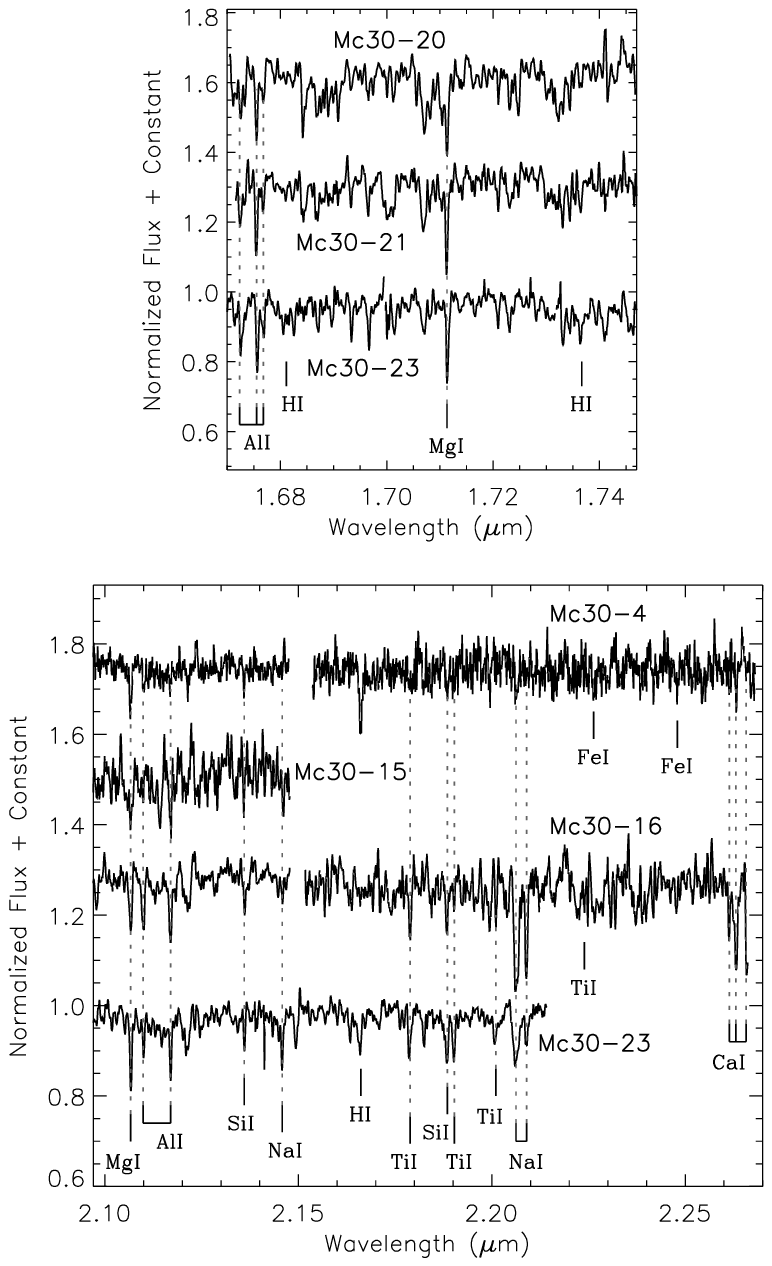}
        \caption{H- and K-band spectra of cool stars in the Mercer 30 field. Mc30-23 is the only object that has been observed in both bands. Unlike Fig. \ref{fig:hotspectra}, spectra have been corrected for radial velocities; vacuum wavelengths are used. 
              }
      \label{fig:coolspectra}
    \end{figure}

The reduced spectra are shown in Figs. \ref{fig:hotspectra} and \ref{fig:coolspectra}, where objects are labeled with the same identification number as appears in Fig. \ref{fig:finders}. To identify spectral features and determine spectral types and luminosity classes, we have relied on the available spectral atlases in the H and K bands \citep{kleinmann-hall86, eenens+96, origlia+93, hanson-conti94, morris+96, wallace-hinkle96, wallace-hinkle97, figer+97, meyer+98, hanson+96, hanson+98, hanson+05, ivanov+04}; wavelengths in vacuum of spectral lines are taken from the Van Hoof's Atomic Line List\footnote{http://www.pa.uky.edu/$\sim$peter/atomic}. The resulting classification is presented in Table \ref{tab:stars}.

Specifically, O and early-B subtypes are based on the \ion{He}{ii}/\ion{He}{i} strength ratio, and luminosity classes are distinguished through the width of the hydrogen lines, as well as the presence of emission components for the most luminous objects. Wolf-Rayet (WR) stars are classified through their characteristic broad emissions and their subtypes are, again, based on the \ion{He}{ii}/\ion{He}{i} ratio. With regard to cool stars, the observed spectral ranges do not cover any luminosity diagnostic (typically, the CO bands beyond $2.29 \mu \mathrm{m}$), therefore we only perform a rough classification of F/G/K/M-types based on the existence of hydrogen lines and the intensity of \ion{Ca}{i} and \ion{Na}{i} lines in the $K$ band.

  \begin{figure}
    \centering
    \includegraphics[height=8cm, bb=0 -19 333 1039, clip]{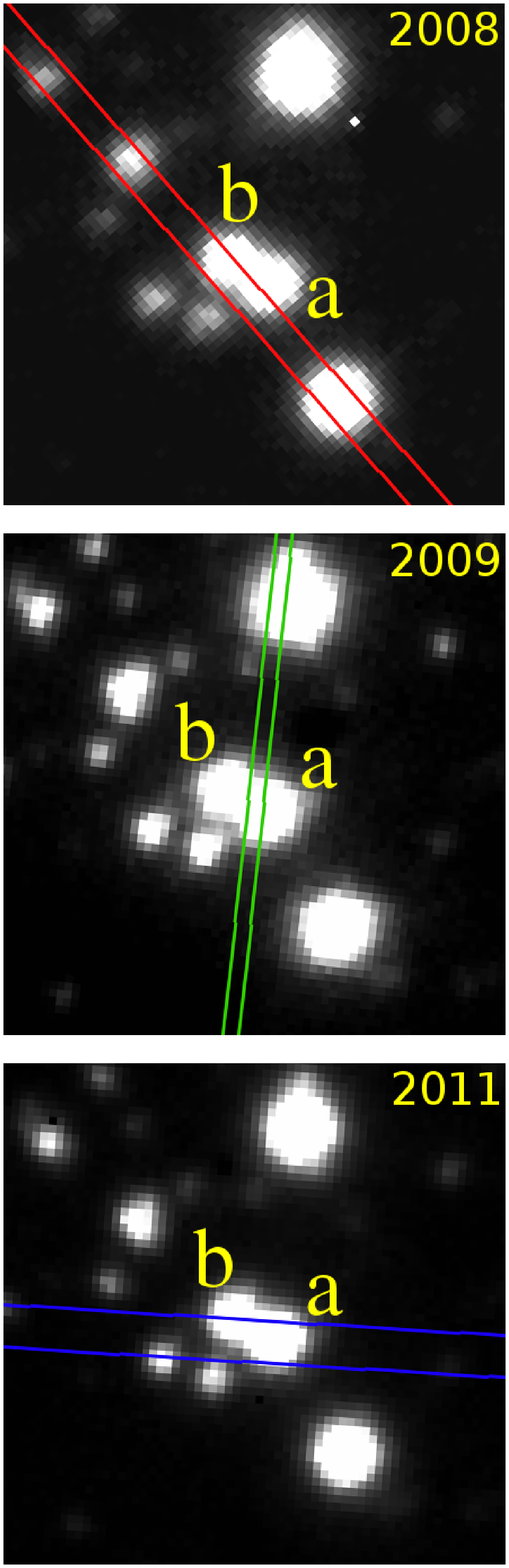}
    \quad
    \includegraphics[height=8cm]{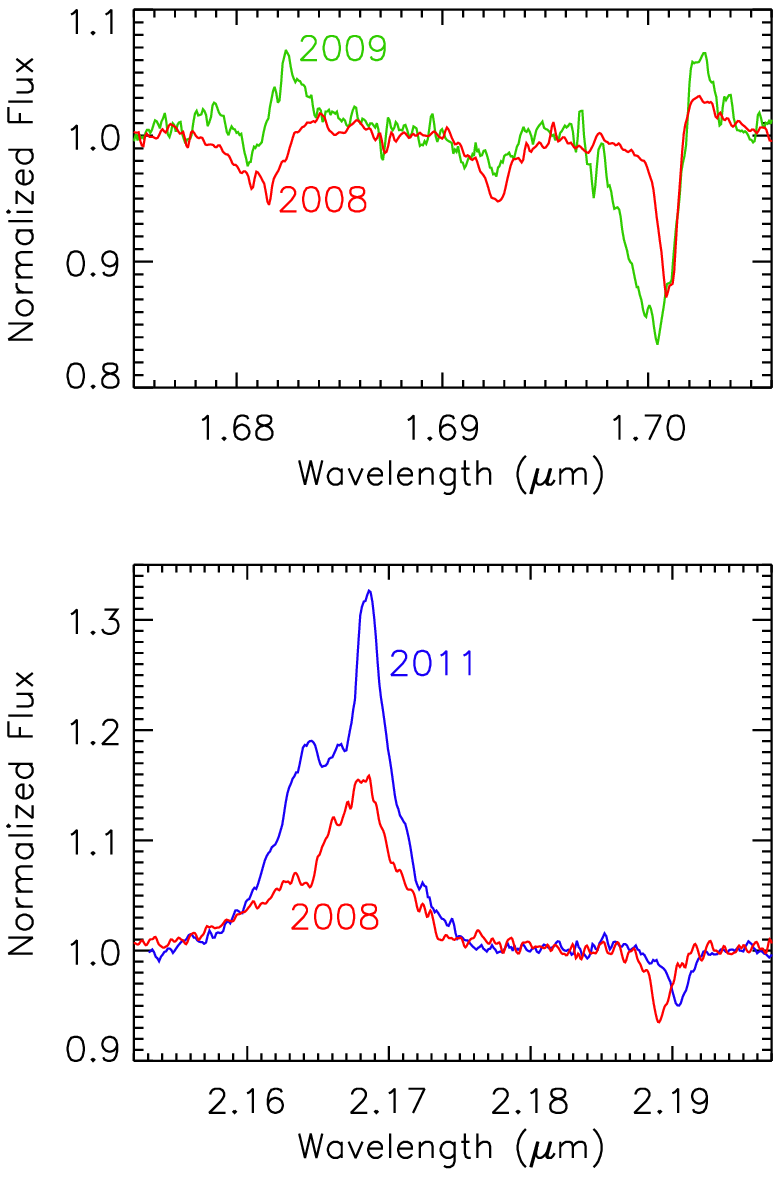}
    \caption{Left: Slit positions used for observing Mc30-6, superimposed on $10'' \times 10''$ cut-outs of the corresponding ISAAC acquisition images, where North is up and East is left. The double object close to the center of each image is Mc30-6, and each component is labeled. Right: comparison of the extracted spectra of Mc30-6 at the ranges where variations are more striking. Each spectrum is labeled with the year of observation. Colors are coded as in Fig. \ref{fig:hotspectra}.}
    \label{fig:slits}
  \end{figure}

A troublesome case of spectral classification is Mc30-6, which is the same object as star \#3 of \citet{kurtev+07}. As shown in Fig. \ref{fig:slits}, this object was observed with three different slit orientations, yielding a significantly varying spectrum. Narrow and broad features that are typical of O supergiants and nitrogen-rich Wolf-Rayet stars (WN), respectively, appear mixed in different proportions at each slit position. On the other hand, the NICMOS/HST K-band image (see Fig. \ref{fig:finders}) is able to resolve \object{Mc30-6} in two closely located point sources we will name the southwestern (and brightest) star \object{Mc30-6a} and the northeast one \object{Mc30-6b}.
Although spectra of both stars appear mutually contaminated in the ISAAC/VLT acquisition images, close inspection of the slit placements (Fig. \ref{fig:slits}) reveals that Mc30-6a is the major contribution in the 2009 and 2011 observations (especially in the H band thanks to the narrower slit) and a minor contaminant of Mc30-6b in the 2008 observation. This enables us to assign an Ofpe/WN9 classification for Mc30-6a \citep[in agreement with][]{kurtev+07}, and O6 If type for Mc30-6b.

The remaining three stars observed by \citet{kurtev+07} in the K band are also included among our spectroscopic targets. Stars \#1, \#2, and \#4 of the cited paper correspond to \object{Mc30-8}, \object{Mc30-7}, and \object{Mc30-1}, respectively. We have derived the same spectral types for these objects except Mc30-1 (= \#4). \citet{kurtev+07} assigned a weak-lined WN classification to this object based on the Br$\gamma$ and \ion{He}{ii} broad emissions. However, our enhanced spectra (higher resolving power and signal-to-noise ratio (S/N), as well as a wavelength coverage that includes the H band), shows narrow absorption components in the hydrogen and helium lines that favour an intermediate-O classification. Additionally, the broad emissions and P-Cygni type profiles hint at an extremely bright supergiant or hypergiant luminosity class (as with \object{Mc30-2}).

\begin{table*}
  \caption{Equatorial coordinates, magnitudes, and spectral types of spectroscopically observed stars in the Mercer 30 field.}             
  \label{tab:stars}      
  \centering          
  \begin{tabular}{c c c c c c c}       
  \hline\hline                          
  ID        & R.A.& Declination &  $H$\tablefootmark{a}  &    $K_S$\tablefootmark{a}    & Source & Spectral type  \\
  \hline
 Mc30-1  &  $12^h 14^m 31.54^s$ &  $-62^\circ 58' 54.3''$ & $ 9.06 \pm 0.01$ & $ 8.54 \pm 0.01$ & HST & O6-7.5 If+\\
 Mc30-2  &  $12^h 14^m 31.64^s$ &  $-62^\circ 58' 48.1''$ & $ 8.63 \pm 0.01$ & $ 8.14 \pm 0.01$ & HST & B1-4 Ia+\\
 Mc30-3  &  $12^h 14^m 31.65^s$ &  $-62^\circ 58' 45.1''$ & $11.62 \pm 0.02$ & $11.09 \pm 0.02$ & HST & O6 If\\
 Mc30-4  &  $12^h 14^m 22.11^s$ &  $-62^\circ 58' 59.9''$ & $10.99 \pm 0.04$ & $10.71 \pm 0.04$ & VVV & F-G\\
 Mc30-5  &  $12^h 14^m 30.82^s$ &  $-62^\circ 58' 53.2''$ & $13.06 \pm 0.03$ & $12.56 \pm 0.04$ & HST & O7-8 IV-V\\
 Mc30-6a &  $12^h 14^m 31.73^s$ &  $-62^\circ 58' 52.1''$ & $ 9.56 \pm 0.01$ & $ 9.02 \pm 0.01$ & HST & Ofpe/WN9 \\
 Mc30-6b &  $12^h 14^m 31.83^s$ &  $-62^\circ 58' 51.8''$ & $10.25 \pm 0.01$ & $ 9.66 \pm 0.01$ & HST & O6 If \\
 Mc30-7  &  $12^h 14^m 33.10^s$ &  $-62^\circ 58' 51.0''$ & $10.25 \pm 0.01$ & $ 9.67 \pm 0.01$ & HST & WN6\\
 Mc30-8  &  $12^h 14^m 33.91^s$ &  $-62^\circ 58' 48.7''$ & $10.30 \pm 0.01$ & $ 9.64 \pm 0.01$ & HST & WN7\\
 Mc30-9  &  $12^h 14^m 33.03^s$ &  $-62^\circ 58' 41.9''$ & $11.25 \pm 0.02$ & $10.68 \pm 0.02$ & HST & O6-7 I-III\\
Mc30-10  &  $12^h 14^m 32.04^s$ &  $-62^\circ 58' 52.2''$ & $11.84 \pm 0.02$ & $11.28 \pm 0.02$ & HST & O4 I-III\\
Mc30-11  &  $12^h 14^m 32.10^s$ &  $-62^\circ 58' 49.6''$ & $11.33 \pm 0.02$ & $10.77 \pm 0.02$ & HST & O5.5-6 I-II\\
Mc30-12  &  $12^h 14^m 32.18^s$ &  $-62^\circ 58' 46.9''$ & $13.17 \pm 0.03$ & $12.66 \pm 0.04$ & HST & O7.5-8.5 III-V\\
Mc30-13  &  $12^h 14^m 32.73^s$ &  $-62^\circ 58' 55.0''$ & $11.63 \pm 0.02$ & $11.13 \pm 0.02$ & HST & O5.5-6 I-II\\
Mc30-14  &  $12^h 14^m 31.33^s$ &  $-62^\circ 58' 42.3''$ & $13.39 \pm 0.03$ & $12.94 \pm 0.05$ & HST & O9-B3 II-V\\
Mc30-15  &  $12^h 14^m 27.46^s$ &  $-62^\circ 58'  6.2''$ & $14.027 \pm 0.006$ & $13.524 \pm 0.002$ & VVV & G-M\\
Mc30-16  &  $12^h 14^m 26.55^s$ &  $-62^\circ 57' 57.7''$ & $13.08 \pm 0.01$ & $12.881 \pm 0.002$ & VVV & K-M\\
Mc30-17  &  $12^h 14^m 32.19^s$ &  $-62^\circ 58' 50.7''$ & $13.00 \pm 0.03$ & $12.46 \pm 0.04$ & HST & O8.5-B1 II-V\\
Mc30-18  &  $12^h 14^m 32.34^s$ &  $-62^\circ 58' 48.1''$ & $11.99 \pm 0.02$ & $11.51 \pm 0.02$ & HST & O7.5-8.5 I-II\\
Mc30-19  &  $12^h 14^m 32.53^s$ &  $-62^\circ 58' 42.8''$ & $11.67 \pm 0.02$ & $11.09 \pm 0.02$ & HST & O6.5-7 I-III\\
Mc30-20  &  $12^h 14^m 30.79^s$ &  $-62^\circ 58' 39.5''$ & $10.42 \pm 0.01$ & $ 9.68 \pm 0.01$ & HST & K-M\\
Mc30-21  &  $12^h 14^m 33.90^s$ &  $-62^\circ 58'  7.9''$ & $12.24 \pm 0.04$ & $11.57 \pm 0.03$ & VVV & K-M\\
Mc30-22  &  $12^h 14^m 31.41^s$ &  $-62^\circ 59'  1.4''$ & $13.36 \pm 0.03$ & $12.87 \pm 0.04$ & HST & O9 III-V\\
Mc30-23  &  $12^h 14^m 29.44^s$ &  $-62^\circ 58' 58.9''$ & $11.42 \pm 0.04$ & $11.08 \pm 0.04$ & VVV & G\\
  \hline
\end{tabular}
\tablefoot{
\tablefoottext{a}{The F160W and F222M magnitudes of HST sources have been converted to $H$ and $K_S$ (see Section \ref{sec:cmd}).}
}
\end{table*}

As shown in Table \ref{tab:stars}, we  identified 18 early-type stars (i.e., OB and WR spectral types) and six late-type stars. While every confirmed hot star is located at angular distances below 15 arcseconds from the cluster center, cool stars are distributed over the outskirts (see Fig. \ref{fig:finders}). This fact hints that late-type stars are probably foreground or background. Moreover, the only cool luminous objects that could be present in YMCs are red supergiants, which would be the brightest cluster members at near-infrared wavelengths \citep[see e.g.,][]{figer+06, davies+07}. Since none of the late-type stars in the Mercer 30 field dominates the infrared light of the cluster (see Fig. \ref{fig:finders}; this will be proven quantitatively with the color-magnitude diagram at Section \ref{sec:cmd}), membership can be discarded for these objects. Additional membership evidence based on radial velocities will be presented in Section \ref{sec:vrbin}

    \begin{figure}
      \centering
        \includegraphics[width=0.9\hsize, bb=15 5 330 380, clip]{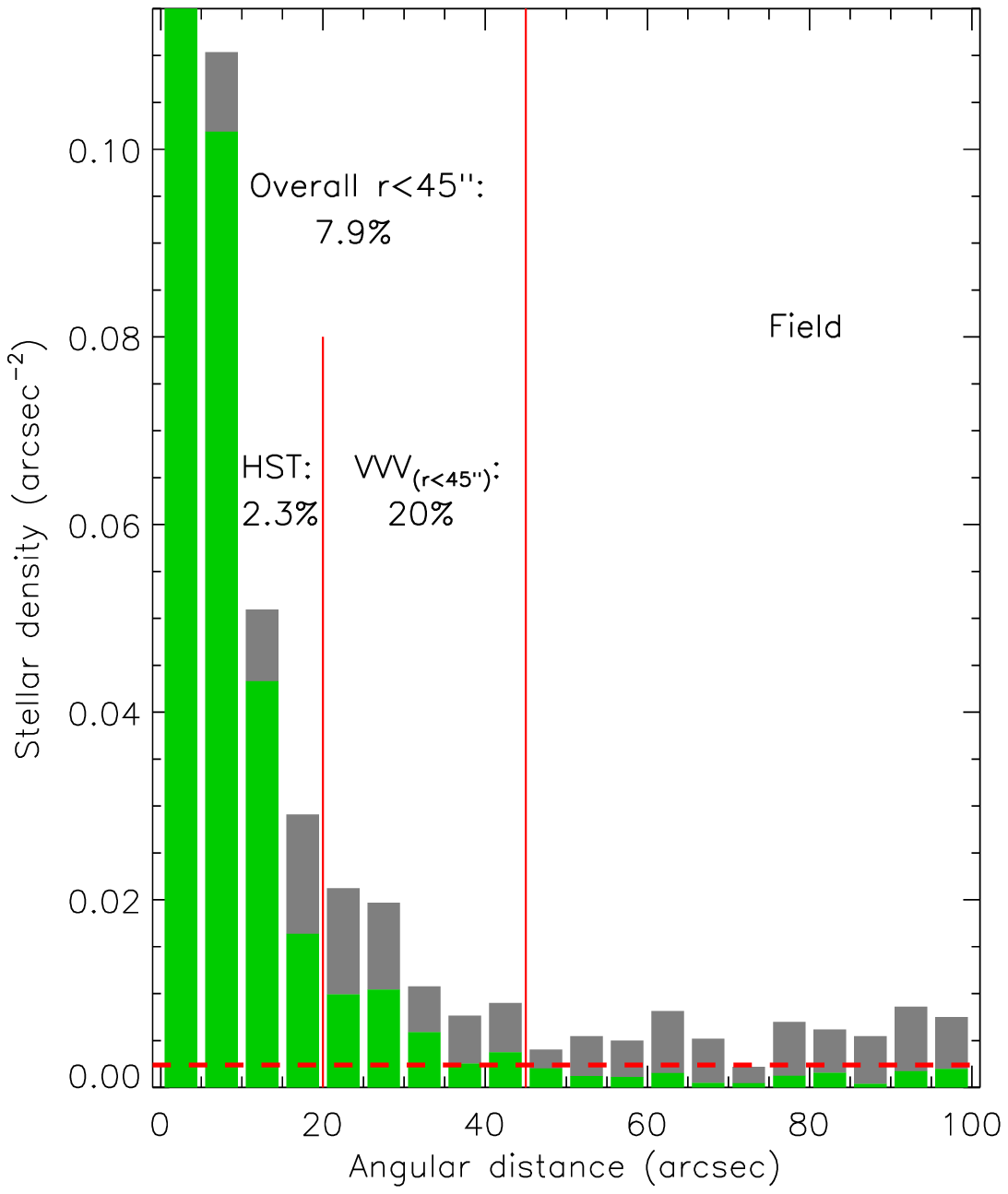}
        \caption{Radial profile of stellar density, centered at R.A. $ = 12^{h} 14^{m} 32.15^{s}$, Dec. $= -62\degr 58' 50.1''$ and binned by 5 arcseconds. Gray bars represent the whole population with $K_S < 15$, and green bars show the fraction of these that fall within the color range of OB cluster members ($0.4 \le H-K_S \le 0.7$). For the sake of clarity, the density values of the first bin (0.2037 and 0.2164) are outside the axis range. The radial cuts for HST and VVV photometry are shown as red vertical lines, and the density limit we use to establish the outer boundary of the cluster is indicated as a red dashed line. For each cluster region, we also provide the probability that a star that fulfills the color cut is actually a contaminant.
              }
      \label{fig:denshisto}
    \end{figure}

  \subsection{Color-magnitude diagram and contaminating sources}
    \label{sec:cmd}

As stated in Section \ref{sec:photometry}, we jointly use  the NICMOS/HST data for $r \leq 20''$ with the VVV photometry outside that angular radius. Since these datasets employ different photometric systems, we apply a transformation between them. \citet{kim+05} obtained a conversion between $K_S$ and $F222M$ that is only valid for the range $0.110 \le F160W-F222M \le 0.344$, but the majority of sources in the Mercer 30 field have significantly redder colors. Therefore, we prefer to build our own photometric transformation using sources in common. To avoid the nonlinear regime of VVV \citep{gonzalez+11, saito+12} and large uncertainties of faint stars, we have only taken H-band magnitudes ranging from 13 to 16.5. Linear fitting of these data yield the following equations:

\begin{eqnarray}
  H & = & F160W - 0.164 \cdot (F160W-F222M) - 0.208\\
  K_S & = & F222M + 0.082 \cdot (F160W-F222M) - 0.191
.\end{eqnarray}

From now on, we will only use $H$ and $K_S$ magnitudes.

    \begin{figure}
      \centering
        \includegraphics[width=0.9\hsize, bb=25 5 325 495,clip]{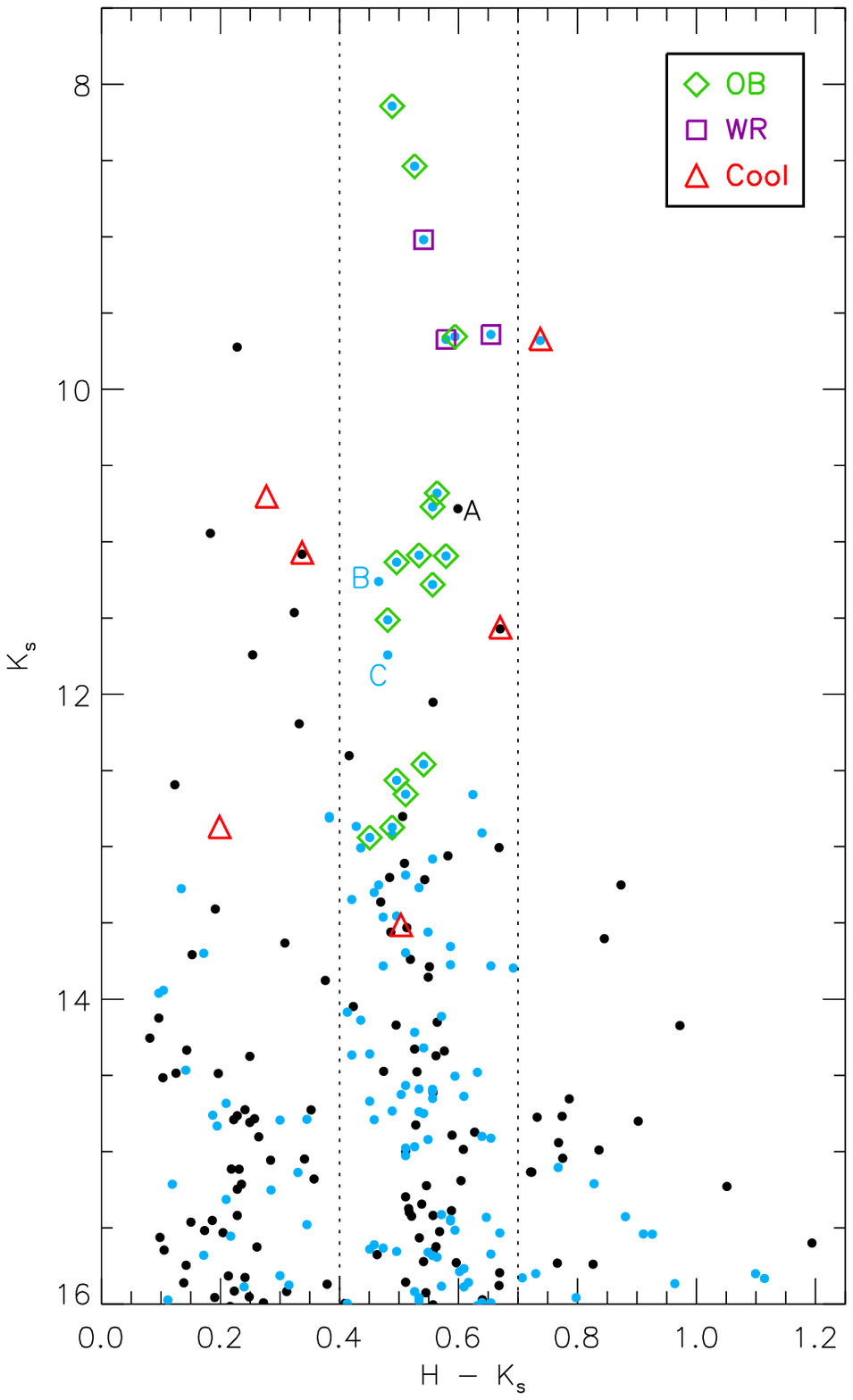}
        \caption{Color-magnitude diagram of Mercer 30 up to a radius of $45''$, with dotted lines enclosing the color range used for selection of hot massive cluster members. Black dots are VVV data and blue dots are the HST photometric measurements converted to the VVV passbands. Spectroscopically observed objects are highlighted with symbols, even if they are located at $r > 45''$. A, B, and C are the three candidate high-luminosity cluster members that will be discussed in Section \ref{sec:flux}.
              }
      \label{fig:cmd}
    \end{figure}

A great majority of the HST sources have colors $0.4 \le H-K_S \le 0.7$, with no exception for $K_S < 12.8$. Also, every spectroscopically confirmed OB or Wolf-Rayet star lies within these color limits. These clues lead us to use this range as a color cut to identify candidate hot massive cluster members (we note, however, that this color cut is not valid for fainter and cooler cluster members, which would show redder colors). We test this criterion, calculating the radial profile of surface density of stars, taking only sources bright enough ($K_S < 15)$ to minimize incompleteness effects. As shown in Fig. \ref{fig:denshisto}, stars within the aforementioned color range are responsible for the sharp density increase towards the cluster center, while the radial profile barely varies if we only consider objects with a redder or bluer color. This fact further supports our color-selection method.

To estimate the ratio of contaminants (i.e., stars that fulfill the color criterion for OB and WR cluster members, despite being unrelated to the cluster), we take the average density of color-selected stars for $50'' < r < 100''$, yielding $\rho^\mathrm{Field}_{[0.4,0.7]} = 0.00119~\rm{arcsec}^{-2}$, (or equivalently, 19\% of all $K_S < 15$ sources in this region). The outer boundary of the cluster is chosen as the radius where the surface density of color-selected stars fall below $2 \cdot \rho^\mathrm{Field}_{[0.4,0.7]}$ (red dashed line in Fig. \ref{fig:denshisto}), which occurs at $r \approx 45''$.
Assuming that the density of contaminants $\rho^\mathrm{Field}_{[0.4,0.7]}$ is constant, we calculate the ratio of contaminants within the cluster area ($r < 45''$) and the subregions of the HST ($r < 20''$) and the VVV photometry ($20'' < r < 45''$), yielding the percentages that are shown in Fig. \ref{fig:denshisto}. In the absence of information about spectral types, these ratios express the probability that a star that fulfills $0.4 \le H-K_S \le 0.7$ is not a cluster member. 

The color-magnitude diagram (CMD) of Mercer 30 up to the $45''$ border is presented in Fig. \ref{fig:cmd}, distinguishing the above-explained HST and VVV regions. While the bulk of the HST sources fulfill the color cut, VVV sources are more scattered over the color range (see also percentages in Fig. \ref{fig:denshisto}). Interestingly, two gaps are apparent in the color-selected region at $K_S  \approx 10$ and $K_S  \approx 12$, especially if we only take the inner 20 arcseconds. These gaps split the spectroscopically observed cluster members into three luminosity groups: hypergiants or luminous supergiants and WR stars (upper part), OB normal supergiants (middle), and OB non-supergiants. Remarkably, 13 out of 17 stars in the first two groups have been spectroscopically classified, which  allows us to reach a nearly complete knowledge of the hot luminous population of Mercer 30.

  \subsection{Modeling}
  \label{sec:modeling}


To obtain accurate physical parameters of cluster members, we have used a large grid of models computed with
the \texttt{CMFGEN} code \citep{hillier90,hillier-miller98}. In a nutshell, \texttt{CMFGEN} solves  
the radiative transfer equation in spherical geometry iteratively for the non-LTE expanding atmospheres of early-type stars.
The model inputs are the 
stellar luminosity, $L_\star$;
the effective temperature, $T_\mathrm{eff}$; the stellar radius, $R_\star$;
the surface gravity, log\,$g$; the mass-loss rate, $\dot M$; 
the wind velocity law, $v(r)$ (which is parametrized by the terminal
value $\varv_\infty$ and the shape exponent $\beta$); 
the microturbulence velocity, $\xi_\mathrm{mic}$; the chemical element abundances at 
the stellar surface, $X_{H}$, $Y_{He}$, $Z_{C}$, $Z_{N}$, etc; and
the clumping law $f_{cl}(r)$.
For the latter, various functional forms have been used \citep{hillier-miller99, najarro+09}.
The code produces a synthetic spectrum that can be shifted and convolved according to other parameters that 
are taken from the instrument configuration or constrained directly from the observed spectra, namely the 
spectral resolution, $R$; the radial velocity, $\varv_\mathrm{LSR}$; the macroturbulence velocity, 
$\zeta_\mathrm{mac}$; and the projected rotational velocity, $\varv \sin i$. 
The stellar luminosity, $L_\star$, is finally calibrated by fitting the intrinsic $(H-K_S)_0$ colors 
(derived from synthetic photometry) to the observations, using the reddening and distance 
results that are calculated in Section \ref{sec:distance}).
We make  use of the transformed radius, $R_T \propto R_\star (\varv_\infty
\sqrt{f_{cl}}/ \dot M)^{2/3}$ \citep{schmutz+89, hillier-miller99}, to
scale $\varv_{\infty}$, $\dot M$ and $R_\star$ from our model to match the
derived luminosity.

Our  model grid for Milky Way massive stars, which is currently composed of more than 2 000 models and is being constantly updated, extensively covers
 the parameter domain encompassing the hot stellar types of Mercer 30 cluster members. 
Specifically, temperature and gravity ranges span from early-O to early-B types for all the expected luminosity 
classes, from dwarfs to hypergiants. On the other hand, models for all the WN subtypes (including hydrogen-rich) 
with very different mass-loss rates are also included, ranging from so-called slash (Of/WN) stars, which overlap with 
the most extreme Of types, to the objects with the most dense winds.
We assume solar metallicity \citep{asplund+09} and allow for different
helium and CNO equilibrium abundances.

  \begin{figure}
      \centering
        \includegraphics[width=0.9\hsize]{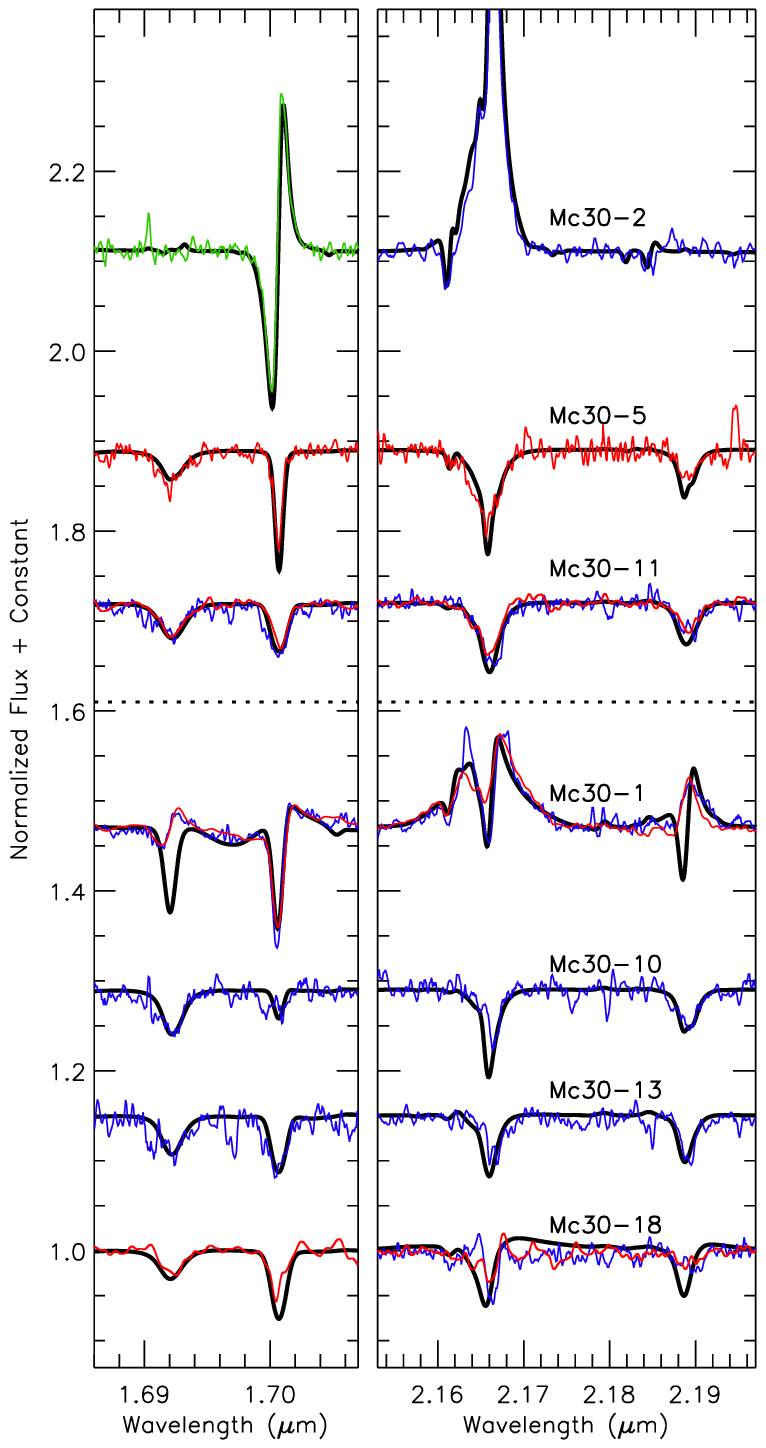}
        \caption{Relevant portions of example models (black lines) fitted to observations (color code of Fig. \ref{fig:hotspectra}). The full plots are shown in Figs. \ref{fig:mod1} and \ref{fig:mod2}. Unlike Fig. \ref{fig:hotspectra}, radial velocities of observed spectra have been corrected with the values of Tables \ref{tab:models} and \ref{tab:vr_variable}, and the higher resolution spectra have been degraded to the lower resolving power when two observations are simultaneously fitted. The observed spectra above the dotted line are well fitted to a single model. Below that line, the spectra show clear signs of binarity (see Section \ref{sec:vrbin}); in these cases, we fit each dominant source.
              }
      \label{fig:modexamp}
  \end{figure}

For each observed early-type star, we  searched the grid for the
best-fitting model. When two different epochs are available for the
same wavelength band, the spectrum with the highest spectral resolution is  degraded
to the lowest resolving power and both spectra are then fitted
simultaneously. We note, however, that there are some cases that display small but significant variations that are associated
with binarity, which will be discussed in Section \ref{sec:vrbin}.
Regarding the special case of Mc30-6, approximate models for stars 6a and 6b are 
fitted to the spectra where the corresponding sources are dominant (see Fig. \ref{fig:slits}).

The search process over the grid was assisted by the known spectral diagnostics in the $H$ and $K$ bands, which 
are partially dependent on stellar types; we explain the main
constraints below. The intensity ratios between
\ion{He}{i} and \ion{He}{ii} lines are optimum $T_\mathrm{eff}$ estimators for O and WR types. If hydrogen lines 
appear clearly  in absorption, gravity can be well determined from
their line profiles shapes, provided that their profiles are purely
photospheric (which is usually valid for OB non-supergiants and even some relatively faint supergiants). 
On the other hand, lines significantly contaminated by dense winds
provide excellent diagnostics to obtain $\dot M$ and $f_{cl}(r)$ and
constrain the velocity field.

When the best-fitting models in the grid required further fine tunning to match the observations, 
additional \texttt{CMFGEN} micro-grids with slight variations in the
stellar parameters were computed. This was especially necessary for carbon
and nitrogen line fitting, since the corresponding abundances seem to be solar and slightly subsolar in
Mercer 30. Further details of the modeling process, focusing on individual stars, are presented in 
Appendix \ref{sec:app}. While an abridgement of model fitting is shown in Fig. \ref{fig:modexamp}, 
the full plots for all modeled stars also appear in Appendix \ref{sec:app}. Spectra of Mc30-12, Mc30-14,
and Mc30-17 are absent since the low S/N and poor spectral coverage impede any acceptable fit to models.

\begin{table*}
  \caption{Main results of \texttt{CMFGEN} modeling of Mercer 30 cluster members.}             
  \label{tab:models}      
  \centering          
  \begin{tabular}{l l l l l l l l l l l l}       
  \hline\hline                          
ID&$T_\mathrm{eff}$~[kK]&$\log g$&$\varv_\mathrm{LSR}~[\mathrm{km~s}^{-1}]$&$X_\mathrm{H}$\tablefootmark{a}&$Y_\mathrm{He}$\tablefootmark{a}&$(H-{K_S})_0$&$R_T$&$M_\star/M_\odot$&$\log (L_\star/L_\odot)$&$\log Q^\mathrm{H}$\\
  \hline
Mc30-1  & 32.2 &3.1\tablefootmark{c}&var. (binary)\tablefootmark{b}& 0.43 & 0.57 &$-$0.047 & 49.4 &99\tablefootmark{c}& 6.51 & 50.15  \\
Mc30-2  & 21.2 &2.4\tablefootmark{c}& 38                           & 0.61 & 0.37 &   0.011 & 79.8 &51\tablefootmark{c}& 6.02 & 48.84  \\
Mc30-3  & 39.3 &  3.80              & 22                           & 0.71 & 0.28 &$-$0.112 & 222.0&  73               & 5.83 & 49.54  \\
Mc30-5  & 36.7 & 3.80      & 20                           & 0.71 & 0.28 &$-$0.110 &172.9&  19       & 5.13 & 48.75  \\
Mc30-6a & 29.9 &3.1\tablefootmark{c}&var. (binary)\tablefootmark{b}& 0.71 & 0.28 &   0.003 & 61.1 &62\tablefootmark{c}& 6.13 & 49.67  \\
Mc30-6b & 30.5 &3.1\tablefootmark{c}& 42                           & 0.71 & 0.28 &$-$0.063 & 61.9 &54\tablefootmark{c}& 6.03 & 49.61  \\
Mc30-7  & 41.4 &3.6\tablefootmark{c}&var. (binary)\tablefootmark{b}& 0.33 & 0.66 &  0.036  & 26.6 &60\tablefootmark{c}& 6.24 & 50.06  \\
Mc30-8  & 38.1 &3.5\tablefootmark{c}& 31                           & 0.42 & 0.57 &   0.097 & 18.5 &49\tablefootmark{c}& 6.07 & 49.86  \\
Mc30-9  & 34.5 &  3.50              &var. (binary)\tablefootmark{b}& 0.71 & 0.28 &$-$0.107 & 189.8&  61               & 5.83 & 49.44  \\
Mc30-10 & 39.5 &  3.65              &var. (binary)\tablefootmark{b}& 0.71 & 0.28 &$-$0.096 & 178.1&  42               & 5.75 & 49.51  \\
Mc30-11 & 36.8 &  3.65              & 35                           & 0.71 & 0.28 &$-$0.102 & 178.1&  73               & 5.87 & 49.53  \\
Mc30-13 & 36.3 &  3.65              &var. (binary)\tablefootmark{b}& 0.71 & 0.28 &$-$0.086 & 189.8&  46               & 5.64 & 49.30  \\
Mc30-18 & 35.5 &  3.65              &var. (binary)\tablefootmark{b}& 0.71 & 0.28 &$-$0.098 & 99.2 &  34               & 5.48 & 49.08  \\
Mc30-19 & 36.0 &  3.50              & 34                           & 0.71 & 0.28 &$-$0.104 & 253.0&  40               & 5.72 & 49.40  \\
Mc30-22 & 32.5 &  3.80              & 25                           & 0.71 & 0.28 &$-$0.111 & 382.8&  17               & 4.87 & 48.19  \\                                             
  \hline
\end{tabular}
\tablefoot{\tablefoottext{a}{Surface abundances are expressed in mass fraction.} \tablefoottext{b}{The varying radial velocities of binary stars are shown in  Table \ref{tab:vr_variable}.} \tablefoottext{c}{Models with dense winds whose surface gravities cannot be reliably inferred; masses have been estimated using the method of \citet{grafener+11} (see text for a detailed description).}
}
\end{table*}

The final values of the main parameters for the best-fitting models are presented in Table \ref{tab:models},
while other results of modeling that are not crucial for our analyses are listed in Appendix \ref{sec:app}. 
Table \ref{tab:models} also shows the intrinsic $(H-K_S)_0$ colors, the stellar masses $M_\star$, and the ionizing fluxes $\log Q^\mathrm{H}$.
For thick wind models, where gravity cannot be directly inferred,  we use the method of \citet{grafener+11} to estimate 
$M_\star$ = $M_\star(L_\star, X_\mathrm{H})$ for very massive stars. These authors provide separate
equations for two extreme cases: chemically homogeneous stars and pure helium cores. Since the chemical 
composition of the Mercer 30 stellar interiors cannot be constrained from our observations, we take the
geometric mean of results from both equations as a very rough estimate for $M_\star$.

Strictly speaking, uncertainties associated with modeling should be calculated on a case-by-case basis, 
testing small variations of each parameter around every position on the parameter space where a final model 
is located \citep[see, e.g.,][]{najarro+09, najarro+11, clark+12} and
may be subject to different degrees of degeneracy depending on the
parameter domain of interest (e.g., $\dot M$ vs $f_{cl}$ and its onset
or $\xi_\mathrm{mic}$).  
Since these topics are beyond the scope of this paper, we only provide here general uncertainty values 
for the fundamental properties, based on our experience in the search process over the grid, which is
subjective to a certain extent. Thus, $T_\mathrm{eff}$ uncertainties are roughly $\sim \pm 1500 \mathrm{K}$
(except for Mc30-22, see \ref{asec-mc30-22}). 
Also, we find that the $\log g$ accuracy is between 0.1 and 0.2, except for objects with very 
dense winds (hypergiants and WR stars), where gravities are unreliable but unimportant for the models.
Conversely, $\dot M$, $\beta$, and $\varv_\infty$ uncertainties are as low as 0.1 dex 
for very thick winds and higher as the winds become thinner. The accuracy of $L_\star$ (and therefore of $R_\star$) 
is dominated by the uncertainty in the distance, which is estimated in Section \ref{sec:distance}. 


  \subsection{Extinction and distance}
  \label{sec:distance}

Interstellar reddening has been estimated using an extinction law of the form $A_\lambda \varpropto \lambda^{-\alpha}$, along with the intrinsic $(H-K_S)_0$ colors of the modeled stars that appear in Table \ref{tab:models}. We assume $\alpha = 1.9,$ based on the exponent values of 1.95, 1.90 recently obtained for the Galactic extinction by \citet{wang-jiang14} using the APOGEE data and an average of previous results, respectively. We obtain an averaged K-band extinction of $\bar A_{K_S} = 0.91 \pm 0.09$ for Mercer 30. The small dispersion on the individually calculated $A_{K_S}$ values (standard deviation: 0.09) implies that there are no noticeable effects of differential extinction. Therefore, the presence of hot massive cluster members outside the color range defined in Section \ref{sec:cmd} and Fig. \ref{fig:cmd} that are due to abnormal local reddening is highly unlikely.

To estimate the spectrophotometric distance to the cluster, we  dereddened each star according to its individual $A_{K_S}$ value, and then we  used the \citet{martins-plez06} calibration of absolute K-band magnitudes for O-type stars. We  discarded Wolf-Rayet stars, as the K-band luminosities of these objects have large dispersions \citep[above 1 magnitude, see][]{crowther+06} even within similar subtypes. The early-type hypergiant stars are also dismissed given that their luminosity dispersion is even higher \citep[see][]{clark+12}.
In summary, we calculated the distances to the modeled OB stars whose luminosity class ranges from main-sequence stars to normal supergiants or, equivalently, modeled stars with $K_S > 10$ (see Table \ref{tab:stars} and Fig. \ref{fig:cmd}), or with luminosities below $10^6 L_\odot$ (see Table \ref{tab:models}). We obtain an average distance of $d = (12.4 \pm 1.7)$ kpc. If stars were dereddened using the above calculated average extinction, $\bar A_{K_S}$ (which also includes WR and hypergiants) instead of each individual $A_{K_S}$, the distance would not change significantly: $d_{\bar A_{K_S}} = (12.6 \pm 1.5)$ kpc.

Comparatively, our distance estimate is significantly higher than the previous result of \citet{kurtev+07}, $d = (7.2 \pm 0.9)$ kpc, which was calculated only using three Wolf-Rayet stars, along with the calibration of WR stars by \citet{crowther+06}. This calibration results in absolute K-band magnitudes of $-4.41$ and $-5.92$ for the WN subtypes in Mercer 30; however, if we took our average extinction and distance results to calculate their actual magnitudes, we would obtain $M_{K_S}^{\mathrm{(Mc30-7)}} = -6.14$ and $M_{K_S}^{\mathrm{(Mc30-8)}} = -6.11$. The fact that these particular WN stars are significantly brighter than the \citet{crowther+06} averages fully explains why \citet{kurtev+07} obtained a much lower distance estimate, as these authors only had available spectroscopic data of Wolf-Rayet stars. Therefore, the distance discrepancy supports our decision of discarding WR stars for spectrophotometric distance calculations.

\begin{table}
  \caption{Radial velocities of cool stars in the Mercer 30 field, and number of lines used for measuring them.}             
  \label{tab:vr_cool}
  \centering          
  \begin{tabular}{c c c c c}
  \hline\hline  
 ID     &$\varv_\mathrm{LSR}[\mathrm{km~s^{-1}}]$&$\sigma_\varv[\mathrm{km~s^{-1}}]$&$N_\mathrm{lines}$\\
\hline  
Mc30-4   &    -86     &      8      &   8  \\
Mc30-15  &     24     &     17      &   3  \\
Mc30-16  &     33     &     11      &  14  \\
Mc30-20  &     13     &      1      &   3  \\
Mc30-21  &    -1      &      9      &   3  \\ 
Mc30-23  &    -14     &     13      &   7  \\ 
  \hline                  
\end{tabular}
\end{table}
  
\subsection{Radial velocities and binary stars}
  \label{sec:vrbin}

Since radial velocities of stars are important to confirm cluster membership, we have tested two different methods that are described below. In any case, the accuracy of these measurements is limited by the wavelength uncertainties associated to our reduction procedures (Section \ref{sec:spectroscopy}), i.e., one tenth of the resolution element.

First, absorption lines were fitted to Gaussian profiles, excluding lines that are significantly blended. Resulting radial velocity measurements for cool stars (Table \ref{tab:vr_cool}) have dispersions that are slightly higher than the wavelength uncertainty. Specifically, spectra with more than three measurements have a median dispersion of $\bar \sigma_\varv = 12~\mathrm{km~s}^{-1}$. However, results for early-type stars are inconsistent, having much higher dispersions, even for lines of the same object that were observed at the same epoch. The problem is due to the ubiquitous wind contamination in spectral features of hot luminous stars, which distorts the profile shapes. The only exceptions are the main-sequence stars \object{Mc30-5} and \object{Mc30-22}, for which we could obtain $\varv_\mathrm{LSR}= 25~\mathrm{km~s}^{-1}$ and $\varv_\mathrm{LSR}= 36~\mathrm{km~s}^{-1}$, respectively.

\begin{table}
  \caption{Modeled cluster members exhibiting significant radial velocity variations, and $\varv_\mathrm{LSR}$ (in km s$^{-1}$).}             
  \label{tab:vr_variable}
  \centering          
  \begin{tabular}{c c c c c}
  \hline\hline
 ID     & \multicolumn{4}{c}{Epoch}   \\
        & 2008   & 2009    & 2011   & 2012 \\ 
  \hline
Mc30-1  &  30    &   35    &  ---   &  55 \\
Mc30-6a &  ---   &   125   &  205   & ---  \\
Mc30-7  & $-$15  &  $-$81  &  110   & ---  \\
Mc30-9  &   34   &   40    &   68   & ---   \\
Mc30-10 &   ---  &   35    &   55   & ---  \\
Mc30-13 &  ---   &    0    &   45   & ---  \\
Mc30-18 & $-$5   &   ---   &   40   & ---   \\
  \hline                  
\end{tabular}
\end{table}

The second method consists of visually finding the velocity shift between models and observations. We test small variations in these velocity shifts to find an uncertainty of $~ 10~\mathrm{km~s}^{-1}$, which is congruent with the aforementioned wavelength accuracy. This also corresponds roughly to the $\varv_\mathrm{LSR}$ differences between both methods for Mc30-5 and Mc30-22; for consistency, we only take the radial velocity values of the model-fitting method for these two stars, as well as for the remaining early-type objects. The results are presented in Table \ref{tab:models}, except when varying velocities (with variations significantly above the expected uncertainty) between different epochs are found (see Table \ref{tab:vr_variable}). The latter are excluded for calculating the average radial velocity of Mercer 30 (based on only early-type stars), $\varv_\mathrm{LSR}^\mathrm{Mc30} = (31 \pm 8)~\mathrm{km~s}^{-1}$. This result will be put in the Galactic context in Section \ref{sec:moving}.

While hot stars with no observed $\varv_\mathrm{LSR}$ variations depart $1.5 \sigma_\varv$, at most, from the average value, the majority of cool stars are outside the $1.5 \sigma_\varv$ range, indicating membership of the former and non-membership of the latter. Although the radial velocities of two late-type objects, Mc30-15 and Mc30-16, would be congruent with the velocity of Mercer 30, their projected distances to the cluster center ($> 50''$) makes membership probabilities extremely unlikely. 

Obviously, the seven objects that are listed in Table \ref{tab:vr_variable} are binary stars. Among them, the four objects that are shown in Fig. \ref{fig:modexamp} below the dotted line have clear evidence of binarity on their spectra, i.e., they have composite features that cannot be well fitted by a single stellar model. The most remarkable case is Mc30-1, which simultaneously displays P-cygni-type and emission profiles in the \ion{He}{ii} and Br-$\gamma$ lines that point to a mid-Of spectral type, together with narrow, strong absorption components in the \ion{He}{i} and H lines that are typical of later subtypes.
Also, multi-epoch composite spectra show significantly variable spectral features, e.g., Br-$\gamma$ in Mc30-18, whose absorption and emission peaks are inverted between the two epochs. Despite the composite features, we have roughly fitted a single stellar model for each object, therefore these models must be interpreted as an approximation for the dominant component. On the other hand, compositeness implies that the secondary component is bright enough to contaminate the spectrum of the primary star. For the S/N of the corresponding observations ($\sim$ 100-200), a K-band luminosity ratio $\log (L_K^{(1)}/L_K^{(2)}) \lesssim 0.5$ is required to display a clearly composite spectrum. As a consequence, the real luminosity of the primary star is somewhat lower than the model luminosity (0.3 dex in the extreme case of $L_\star^{(1)}=L_\star^{(2)}$)

\subsection{Proper motions}

Proper motions for Mercer 30 were produced as part of a large-scale campaign to produce multi-epoch proper motion and parallax catalogues using the VVV $K_S$ filter data, details of which are given in \citet{smith15}. Briefly, we used $K_S$ bandpass pawprint catalogues (with seeing $<1.2$\arcsec) of VVV tile d041 that was available to us as of April 30, 2014. Fitting of the proper motion for each source was performed in \texttt{MATLAB} using a robust technique that involved an iterative reweighting of data points as a function of their residuals. Proper motions for sources detected in up to six pawprints are combined using an inverse variance-weighted average.

We  took the proper motions of all objects detected in our HST and VVV photometric sources within a 1.2 arcmin radius. By means of Gaussian fitting in both the RA and Dec directions, we find that the overall proper motion distribution is centered at $\mu_\alpha \cos \delta = (1.42 \pm 0.16)~\mathrm{mas~yr}^{-1}$; $\mu_\delta = (0.67 \pm 0.13)~\mathrm{mas~yr}^{-1}$. Eight early-type spectroscopic targets (specifically Mc30-1/2/5/6a/6b/10/12/17) could not be measured owing to saturation or contamination in the VVV images. The distribution of proper motions for the remaining ten OB/WR stars is centered at $\mu_\alpha \cos \delta = (1.44 \pm 0.98)~\mathrm{mas~yr}^{-1}$; $\mu_\delta = (0.83 \pm 0.52)~\mathrm{mas~yr}^{-1}$. The corresponding results for the comparison field are $\mu_\alpha \cos \delta = (-0.28 \pm 0.20)~\mathrm{mas~yr}^{-1}$; $\mu_\delta = (-0.08 \pm 0.31)~\mathrm{mas~yr}^{-1}$.

However, these numbers must be interpreted with caution, since differences between the three results are smaller than uncertainties of individual stars. Specifically, uncertainties of $\mu_\alpha \cos \delta$ and $\mu_\delta$ for stars in the $11 < K_S < 15$ range have a median of $2.3 ~\mathrm{mas~yr}^{-1}$ for the whole Mercer 30 field. These median uncertainties increase to $4.2 ~\mathrm{mas~yr}^{-1}$ if we only consider the cluster center ($r < 15''$, where all the confirmed early-type stars are located). Also, the apparent proper motion dispersion is dominated by uncertainties. Therefore, we can neither separate cluster members from field stars nor obtain a noteworthy outcome about kinematics of Mercer 30. We conclude that the long distance to Mercer 30 makes  these tasks unfeasible using our current proper motion data.
  
    \begin{figure}
      \centering
        \includegraphics[width=0.9\hsize, bb=25 7 352 437, clip]{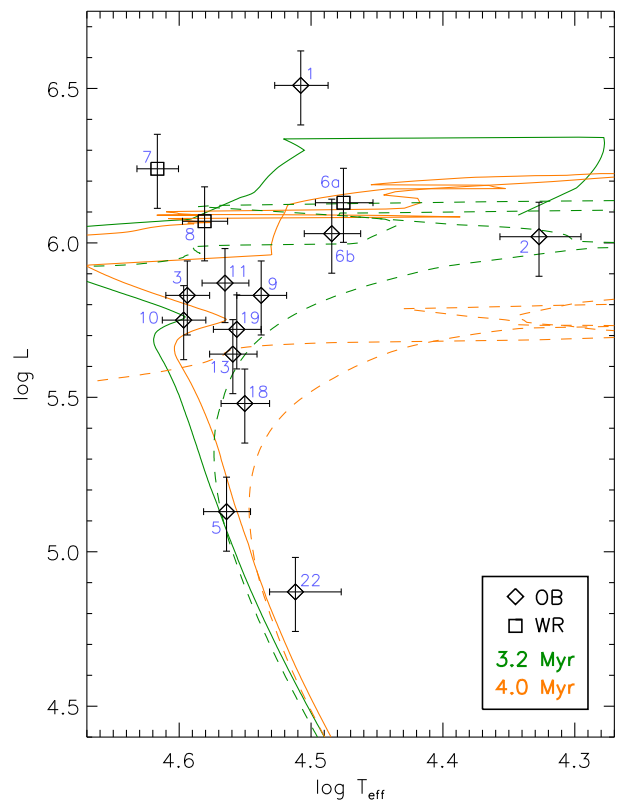}
        \caption{HR diagram of the modeled cluster members showing the best-fit Geneva isochrones for rotating (solid lines) and non-rotating (dashed) stars. The error bars correspond to 1500 K for $T_\mathrm{eff}$, and to the distance uncertainty of $\pm 1.7~\mathrm{kpc}$ for $L_\star$ 
              }
      \label{fig:hrd}
    \end{figure}

  \subsection{HR diagram and age}
    \label{sec:age}

Our nearly complete coverage of the brightest ($K_S < 12$) cluster members in terms of spectroscopy and modeling is the most accurate available tool for finding out the evolutionary state of Mercer 30. At this point, it should be noted that luminosity classes are not good indicators for the evolutionary phase of very massive stars \citep[e.g., a $60 M_\odot$ star may display a supergiant appearance throughout its main-sequence evolution;][]{groh+14}, therefore comparison with stellar evolutionary models is preferrable. We rely on the most recent isochrones at solar metalicity that were produced by the Geneva group \citep{ekstrom+12, georgy+12} for rotating and non-rotating evolutionary models of single stars.

Figure \ref{fig:hrd} shows the Hertzsprung-Russell (HR) diagram of modeled stars, where isochrones for the two youngest ages (3.2 and 4.0 Myr) are also plotted. Subsequent isochrones (5.0 Myr and older) never reach luminosities beyond $10^6 L_\odot$ (since the most massive objects have already exploded), making them incompatible with the most luminous cluster members. All the data points are well fitted by the 3.2 and 4.0 Myr isochrones except Mc30-1. Owing to its binary nature, however, the actual luminosity of the primary component is lower by up to 0.3 dex, as argued in Section \ref{sec:vrbin}.
On the other hand, Mc30-18 and Mc30-22 are located in the concave side (i.e., with lower effective temperature) of the turn-off bend; positions and error bars of these objects put a lower limit on the age as they are only marginally consistent with the 3.2 Myr isochrones. Therefore, Mc30-18 and Mc30-22 provide a lower limit of 3.2 Myr for the age of Mercer 30. Since the upper limit must be strictly lower than 5.0 Myr, we set the age of Mercer 30  as $(4.0 \pm 0.8)~\mathrm{Myr}$., which is consistent with the previous result of \citet{kurtev+07}.

On the other hand, the $K_S \approx 12$ gap in Fig. \ref{fig:cmd} that separates the supergiant and non-supergiant spectral types (see Section \ref{sec:cmd}) roughly corresponds to a luminosity $\log (L/L_\odot) \approx 5.3$ in the HR diagram, at the upper main sequence. We have to clarify here that such a magnitude-luminosity equivalence can only be  set in regions of the HR diagram where objects have roughly the same $T_\mathrm{eff}$. In subsequent sections, this value will be used as a lower cut for the luminosity range where almost all of the cluster members have been spectroscopically observed and modeled; this range includes all the post-main-sequence stars as well as objects near the turn-off point.

\subsection{Ionizing flux}
  \label{sec:flux}

In this section, we calculate the total number of Lyman-continuum photons per second that are emmitted by Mercer 30. The addition of the ionizing fluxes for the modeled stars (Table \ref{tab:models}, last column) is $6.16 \times 10^{50} \mathrm{s}^{-1}$. We note that binarity does not affect this result (despite the models being for single stars), given that luminosities are calibrated with the integrated light of stars independently of their single/binary nature, and $\log Q^\mathrm{H}$ is approximately linear with $\log L_\star$.

This result must be already close to the final value for the whole cluster, given that we have modeled nearly all the high-luminosity cluster members ($K_S < 12$), which are expected to provide the bulk of the ionizing radiation. However, a more accurate calculation that is based on all the possible ionizing sources is required since the outcome will be crucial for the next sections. To include all the non-modeled ionizing sources of Mercer 30, we  approximate their Lyman-continuum flux by means of their modeled photometric analogs (i.e., the closest sources in the CMD, Fig. \ref{fig:cmd}), taking only those objects that fulfill the color criterion, $0.4 \le H-K_S \le 0.7$.

As shown in Fig. \ref{fig:cmd}, only three color-selected objects with $K_S < 12$, labeled as A,B,C, have unknown spectral types. Objects B and C are very close to the cluster center (both at $r < 10''$), which makes it extremely unlikely that they were contaminants (see Fig. \ref{fig:denshisto}). Their $H-K_S$ colors are very similar to \object{Mc30-13} and \object{Mc30-18}, and slightly ($\sim 0.2$ mag) fainter, therefore we assume for these objects $\log Q_B^\mathrm{H} \approx 49.2$ and $\log Q_C^\mathrm{H} \approx 49.0$.
Likewise, object A is photometrically very similar to \object{Mc30-9} and \object{Mc30-11}, and we would assume $\log Q_A^\mathrm{H} \approx 49.5$ if it was a cluster member, however membership is under suspiction for this particular case. Its location in the VVV photometric region ($20'' < r < 45''$) requires that the membership probability is only 80\% (see Fig. \ref{fig:denshisto}). Unfortunately, this is the brightest object with no available spectrum, implying that its hypothetical ionizing flux would introduce a significant uncertainty. Since we aim at calculating a lower limit for the ionizing power of Mercer 30, we  ignore this dubious source.

The five spectroscopically observed stars with $K_S > 12$ which are expected to be among the brightest main-sequence objects, have K-band magnitudes $13 > K_S > 12$. Within this magnitude range, a total of 12 objects fulfill the color-selection criterion (Fig. \ref{fig:cmd}). Taking into account the 7.9\% of contaminants (Fig.\ref{fig:denshisto}), we obtain 11 cluster members in the upper portion of the main sequence that roughly corresponds to the luminosity range of O-type dwarfs, as inferred from known spectral types (Table \ref{tab:stars}).
Mc30-5 and Mc30-22, which are the only modeled objects in this range, can be considered as representative of this O-type dwarf population, based on their positions in the CMD (Fig. \ref{fig:cmd}) and the HR diagram (Fig. \ref{fig:hrd}). To obtain a rough estimate of the contribution of this group of stars, we use an intermediate value between ionizing fluxes of Mc30-5 and Mc30-22, $\log Q^H \approx 48.5$, as the approximate ionizing flux of the remaining nine objects. As a result, the total ionizing flux of the nine brightest non-modeled O-type dwarfs is $2.85 \times 10^{49} \mathrm{s}^{-1}$.

Adding all the above considered contributions, we obtain a lower limit of $6.70 \times 10^{50} \mathrm{s}^{-1}$ for the cluster, and this value would increase very slightly ($7.02 \times 10^{50} \mathrm{s}^{-1}$) if object A was finally confirmed as a cluster member. As the eleven brightest main-sequence stars with $K_S > 12$ only provide 5\% of the total, the contribution of fainter (and cooler) main-sequence objects is assumed to be negligible.

\subsection{Mass}

We take advantage of our detailed knowledge of the post-main-sequence population of Mercer 30 to sample the initial mass function (IMF) in a certain mass range, by counting stars above the luminosity cut, $\log (L_\mathrm{cut}/L_\odot) \approx 5.3$. As we argued in Section \ref{sec:age}, this value is equivalent to an initial mass cut above which nearly all the cluster members are spectroscopically confirmed. Therefore, we need to estimate this mass cut, as well as the initial mass of a hypothetical star that is reaching the supernova (SN) event at the age of the cluster.

Inspection of the Geneva isochrones \citep{ekstrom+12} yields $M_\mathrm{ini}^\mathrm{cut} \approx 31 M_\odot$ for the 4 Myr isochrone and $M_\mathrm{ini}^\mathrm{cut} \approx 33 M_\odot$ for 3.2 Myr, independently of rotation. Therefore we establish $M_\mathrm{ini}^\mathrm{cut} \approx (31 \pm 2) M_\odot$ to account for the age uncertainty. The spectroscopic masses of modeled stars immediately below (Mc30-5, $19 M_\odot$) and above (Mc30-18, $34 M_\odot$) the luminosity cut are consistent with this result, since differences between current and initial masses are still small ($\sim 1 M_\odot$) at the upper main sequence.

The high-mass limit is harder to calculate, since the non-rotating and rotating isochrones lead to significantly different values owing to the slower evolution of rotating stars. Specifically, the 4.0 Myr isochrones end at $M_\mathrm{ini}^\mathrm{SN} \approx 60 M_\odot$ (no rotation) and $M_\mathrm{ini}^\mathrm{SN} \approx 90 M_\odot$ (rotation).  As a compromise, we simply take an intermediate value between the two extreme theoretical values presented above, $M_\mathrm{ini}^\mathrm{SN} \approx (75 \pm 10) M_\odot$. Unlike $M_\mathrm{ini}^\mathrm{cut}$, the $M_\mathrm{ini}^\mathrm{SN}$ result cannot be tested through observational constraints, owing to the following sources of uncertainty.
First, very massive, evolved stars have denser winds and, therefore, less reliable mass estimates, as argued in Section \ref{sec:modeling}. Second, comparisons between initial and spectroscopic mass for the most evolved stars are challenging, since mass loss is highly uncertain and its implementation may vary in different evolutionary models. Moreover, the highest modeled mass ($M_\star^\mathrm{Mc30-1} \approx 99 M_\odot$) actually correspond to a binary system, and other very high values among the most massive stars might have been increased through binary evolution \citep{schneider+14}.

Objects above the luminosity cut are counted following the discussion of Section \ref{sec:flux}, i.e., there are 13 modeled objects plus three high-luminosity cluster member candidates. As discussed in section \ref{sec:vrbin}, several of these 16 objects are confirmed binary stars; taking each binary into consideration for star counts will depend on our knowledge of the secondary star. First, an object whose spectrum is composed of two early-type components must fulfill $\log (L_1/L_2) \lesssim 0.5$, as inferred in Section \ref{sec:vrbin}. Hence, we can ensure that both components are more luminous than $L_\mathrm{cut}$ when the integrated light is at least 0.5 dex more luminous than this value; the only cluster member fulfilling these requirements is Mc30-1, which will be counted as two stars that exceed $M_\mathrm{ini}^\mathrm{cut}$.
The remaining three composite spectra will compute as an uncertain number between 3 and 6, since the secondary objects can be either above or below the luminosity cut. The same conclusion applies to the three confirmed binary systems whose spectrum shows no signs of the companions, therefore these will also be considered as a number of stars between 3 and 6. The remaining six modeled stars will, of course, be counted as single stars. Regarding the three unobserved high-luminosity cluster member candidates, their single/binary status is unknown, and only two of them (B and C) have a very high membership probability, therefore these objects contribute between 2 and 5 stars. As a result, the total number of stars with $L_\star > L_\mathrm{cut}$ can be expressed as $[16, 25]$, or equivalently, $N_\mathrm{luminous} = 20.5 \pm 4.5$.

The IMF functional form of \citet{chabrier05} is calibrated with the above presented results and integrated over the 0.5 - 150 $M_\odot$ range. Although we have assumed the 150 $M_\odot$ limit observationally inferred by \citet{figer05}, we have to remark that this upper limit has been challenged by \citet{crowther+10}\footnote{However, the high-mass values presented by \citet{crowther+10} have recently been  revised downwards (Rubio-D\'iez et al., in prep.).}. Thus, we obtain a total cluster mass of $1.6 \times 10^4 M_\odot$.

This result is affected by two kinds of errors that must be taken into account. On the one hand, the observational uncertainty is found by varying the $M_\mathrm{ini}^\mathrm{cut}$, $M_\mathrm{ini}^\mathrm{SN}$, $N_\mathrm{luminous}$ values within their above calculated uncertainty ranges, yielding $\pm 0.5 \times 10^4 M_\odot$. On the other hand, the discrete nature of the IMF sampling for finite stellar populations leads to stochastic fluctuations \citep[see, e.g.,][]{barbaro-bertelli77,cervino+00,cervino+02}, causing additional errors in the number of inferred stars. To compute this statistical uncertainty, we  performed a Monte Carlo simulation of $10^6$ clusters whose masses are uniformly distributed in the range $3.7 < \log M_\mathrm{cl} < 4.7$, all of them following the \citet{chabrier05} IMF.
For each synthetic cluster, we count the derived number $N_s$ of stars that fulfill $31 M_\odot < M < 75 M_\odot$, and then we select a total of 42344 simulated clusters with $N_s = 20$ or $N_s=21$ (i.e. taking $N_s \approx N_\mathrm{luminous}$). Masses of these selected clusters have an average of $1.6 \times 10^4 M_\odot$ (as expected) and a standard deviation of $0.4 \times 10^4 M_\odot$; the latter is interpreted as the statistical uncertainty associated with this method. After combining the two kinds of errors, we express the final result for the Mercer 30 mass as $(1.6 \pm 0.6) \times 10^4 M_\odot$.

\section{Mercer 30 and the Dragonfish complex}
 
 \subsection{Mercer 30 as part of a moving group}
  \label{sec:moving}

The existence of a $8 \mu \mathrm{m}$ bubble around Mercer 30 (see Fig. \ref{fig:cavities}) suggests a physical association with the Dragonfish star-forming complex. However, securing membership is crucial to enable the application of Mercer 30 results to the whole complex. Hence, we present additional evidence that is based on kinematics below.

 \begin{table}
  \caption{Radial velocity measurements of \ion{H}{II} regions, giant molecular clouds, and masers in the Dragonfish nebula, sorted by angular distance to Mercer 30.}             
  \label{tab:dragonfish_vr}      
  \centering          
  \begin{tabular}{c c c c c c}       
  \hline\hline 
$D_{Mc30}$  & $l$ & $b$ &  $\varv_\mathrm{LSR}$   &    Type     &   Refs.        \\
(arcmin) & ($\degr$) & ($\degr$) & (km s$^{-1}$) &      &       \\  
  \hline
 6.4  &  298.858 &$-$0.436 &  28.5   &    H2R    &  2, 10   \\
 6.9  &  298.868 &$-$0.432 &  25     &    H2R    &    3 10    \\
 7.0  &  298.800 &$-$0.300 &  25     &    H2R      &  11, 10   \\
 7.9  &  298.632 &$-$0.362 &  41     &    MM    &  8, 10   \\
 8.7  &  298.900 &$-$0.400 &  24.2   &     H2R     &  11, 10   \\
 8.7  &  298.90  & $-$0.40  &   29    &     H2R     &  9        \\
 8.7  &  298.900 &$-$0.400 &  25     &     H2R     &  6, 10   \\
12.8  &  298.800 &$-$0.200 &  25     &     GMC     &  10        \\
19.4  &  298.723 &$-$0.086 &  19.5   &    MM    &  8, 10  \\
21.2  &  298.559 &$-$0.114 &  23     &     H2R     &  3, 10   \\
31.9  &  298.228 &$-$0.331 &  31     &     H2R     &  3, 10   \\
31.9  &  298.600 & +0.100 & $-$35    &     GMC     &  10        \\
32.1  &  298.224 &$-$0.341 &   28.5  &    WM       &  4    \\
32.1  &  298.224 &$-$0.342 &    24   &    WM       &  1   \\
32.4  &  298.838 & +0.125 &  21.5   &     H2R     &  2, 10  \\
32.7  &  298.213 &$-$0.343 &  35.2   &    MM    &  8, 10  \\
32.8  &  298.213 &$-$0.338 &    37   &    MM       &  5    \\
33.9  &  298.200 &$-$0.300 &  30.6   &     H2R     &  11, 10  \\
33.9  &  298.20  & $-$0.30  &   34.5  &     H2R     &  9       \\
34.6  &  299.152 & +0.009 &  1.9    &     H2R     &  2, 10  \\
35.7  &  299.013 & +0.128 &  18.8   &    MM    &  8, 10  \\
35.8  &  299.013 & +0.130 &     26   &    WM       &  4    \\
35.8  &  299.015 & +0.129 &     18   &    MM       &  5    \\
36.9  &  299.016 & +0.148 &  23     &     H2R     &  3, 10   \\
36.9  &  299.016 & +0.148 &     25   &    WM     &  1   \\
37.6  &  299.363 &$-$0.257 & $-$37    &     H2R     &  3, 10   \\
39.3  &  299.400 &$-$0.300 & $-$52    &     H2R     &  7, 10   \\
40.8  &  298.187 &$-$0.782 &  16     &     H2R     &  3, 10   \\
41.1  &  298.183 &$-$0.786 &  22.7  &     H2R     &  2, 10   \\
41.7  &  298.177 &$-$0.795 &  25.2  &     MM    &  8, 10  \\
42.9  &  299.400 &$-$0.100 & $-$6     &     GMC     &  10         \\
73.9  &  297.660 &$-$0.973 &    24   &     WM      &  4    \\
74.1  &  297.658 &$-$0.975 &  30.7  &     H2R     &  2, 10   \\
74.3  &  297.655 &$-$0.977 &  26     &     H2R     &  3, 10    \\
77.9  &  297.506 &$-$0.765 &  23     &    H2R     &  3, 10    \\
81.5  &  297.400 &$-$0.500 &  22     &     GMC     &  10         \\
81.9  &  297.406 &$-$0.622 &  27     &    MM    &  8, 10  \\
  \hline                  
\end{tabular}
\tablefoot{The following abbreviations are used. H2R: \ion{H}{II} region; GMC: giant molecular cloud; MM: methanol maser; WM: water maser.}
\tablebib{(1) \citet{braz-epchtein83}; (2) \citet{bronfman+96}; (3) \citet{caswell-haynes87}; (4) \citet{caswell+89}; (5) \citet{caswell+95}; (6) \citet{churchwell+74}; (7) \citet{dickel+72}; (8) \citet{green+12}; (9) \citet{gillespie+77}; (10) \citet{hou-han14}; (11) \citet{wilson+70}.
}
\end{table}

\begin{figure}
      \centering
        \includegraphics[width=\hsize, bb=16 8 380 325, clip]{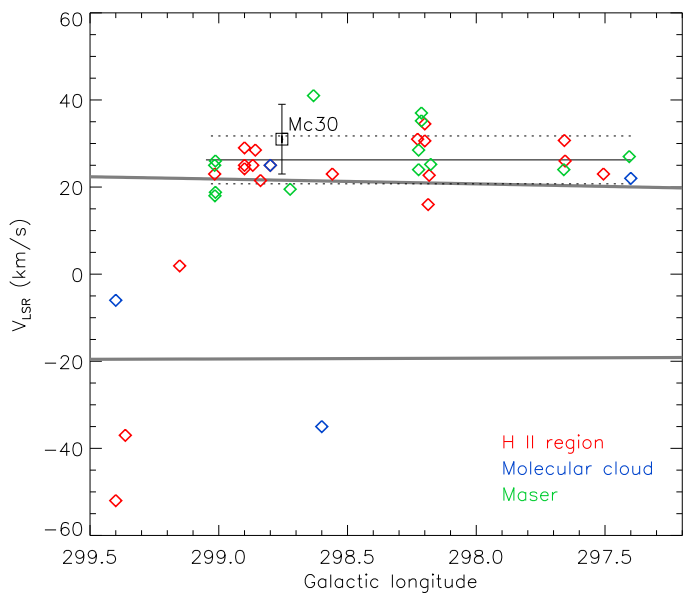}
        \caption{Galactic longitude-velocity diagram of tracers of star formation in the Dragonfish complex that are listed in Table \ref{tab:dragonfish_vr}, as well as Mercer 30. The black thin horizontal lines show the average (continuous line) and standard deviation (dotted) of tracers with  $\varv_\mathrm{LSR} > 10~\mathrm{km~s^{-1}}$. The thick gray curves are the \citet{hou-han14} fit for the Sagittarius-Carina spiral arm, which crosses the diagram twice, at $\sim -20~\mathrm{km~s^{-1}}$ (near part) and at $\sim +22~\mathrm{km~s^{-1}}$ (far part).
              }
      \label{fig:dragonfish_vgal}
    \end{figure}

We have carried out a literature search of radial velocity measurements of objects that are typically associated with star-forming regions, namely \ion{H}{ii} regions, molecular clouds, and different kinds of maser (methanol, water, hydroxyl) in the area covered by the Dragonfish complex. The resulting tracers of star formation and their velocities in the LSR reference frame are presented in Table \ref{tab:dragonfish_vr} and drawn in Fig.  \ref{fig:dragonfish_vgal}. The bulk of velocity measurements is in the range [16,41] km/s; we take the average, $\bar \varv_\mathrm{LSR} = (26.3 \pm 5.5)~\mathrm{km~s^{-1}}$ as the radial velocity value of the Dragonfish complex.
The five radial velocity measurements that are not included in that range are clear outliers ($> 4 \sigma $). Four of them are located in the eastern side of the Dragonfish Nebula ($l > 299.1\degr$), which include the already known foreground \ion{H}{ii} region RCW 64. The fact that no measurement at the eastern Dragonfish region is compatible with the mean velocity indicates that a significant fraction of the emission at  $l > 299.1\degr$ (not only the RCW 64 clump) is physically unrelated to the rest of the Dragonfish  nebula.
  
Fig. \ref{fig:dragonfish_vgal} also shows the kinematical path of the Sagittarius-Carina spiral arm, using the polynomial-logarithmic model with $R^\mathrm{(GC)}_\odot = 8.5$ kpc that \citet{hou-han14} fitted to \ion{H}{ii} regions. The radial velocity of the Dragonfish complex is consistent with being part of the far side of the Saggitarius-Carina arm, although a velocity difference of $\sim +5~\mathrm{km~s^{-1}}$ seems to exist. This peculiar velocity does not pose a problem, given that spiral arm models and theoretical rotation curves are just idealizations that do not necessarily account for kinematical inhomogeneities or irregularities such as bumps and spurs \citep[see, e.g.,][]{alvarez+90, shetty-ostriker06}. We note that \citet{alvarez+90} found a similar anomaly in the velocity field of the Sagittarius-Carina arm at a galactocentric radii of 0.8 - 0.9 $R^\mathrm{(GC)}_\odot$.

We also show in Fig. \ref{fig:dragonfish_vgal} that our radial velocity estimate of Mercer 30 is compatible with the Dragonfish nebula kinematics within uncertainties. Interestingly, the inclusion of the cluster in the Dragonfish complex with its kinematic peculiarity enables us to place Mercer 30 in the Sagittarius-Carina spiral arm.
  
    \begin{figure}
      \centering
        \includegraphics[width=0.8\hsize, bb=15 6 270 242, clip]{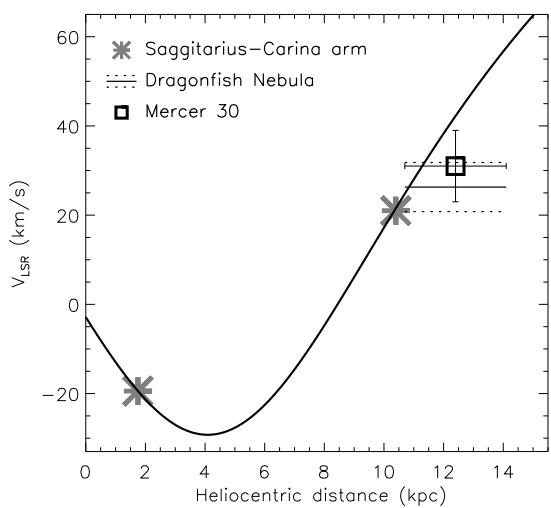}
        \caption{Velocity-distance diagram along the line of sight of Mercer 30, showing the Galactic rotation curve from \citet{brand-blitz93} and its crossing points with the \citet{hou-han14} fit for the Saggitarius-Carina spiral arm. The error bars of Mercer 30 corresponds to the standard deviations of the radial velocity and the spectrophotometric distance.
              }
      \label{fig:mc30_vgal}
    \end{figure}

An additional view of kinematics is provided by the distance-velocity diagram in Fig. \ref{fig:mc30_vgal}, where the Galactic rotation curve of \citet{brand-blitz93} towards Mercer 30 is drawn. The far crossing point of the Saggitarius-Carina arm at 10.4 kpc is slightly less distant than the lower limit of our spectrophotometric estimate for Mercer 30. As with radial velocity, a slight distance difference could be explained in terms of irregularities in the spiral arm. Taking the membership of Mercer 30 to the Dragonfish complex into
account, our combined results of velocity and distance encourage us to interpret the Dragonfish nebula as a star-forming feature in the outer edge (i.e. the convex side) of the Saggitarius-Carina spiral arm.
Consequently, we suggest a heliocentric distance of ($12.4 \pm 1.7$) kpc for the whole star-forming complex based on the Mercer 30 estimate, instead of the $10.8^{+0.6}_{-0.5}$ kpc value that would be obtained if the velocity of the Dragonfish velocity was fitted to the \citet{brand-blitz93} curve (we note, however, that both uncertainty ranges overlap). We also indicate that our distance estimate implies a galactocentric radius of $R^\mathrm{(GC)}_\mathrm{Dragonfish} \approx R^\mathrm{(GC)}_\mathrm{Mc30} = (11.2^{+1.3}_{-1.2} \mathrm{kpc})$, assuming $R^\mathrm{(GC)}_\odot = 8.5 \mathrm{kpc}$.

  \subsection{On the clustered star population}
    \label{sec:clustered}

\begin{figure*}
      \centering
        \includegraphics[width=0.95\hsize, bb=28 15 845 610, clip]{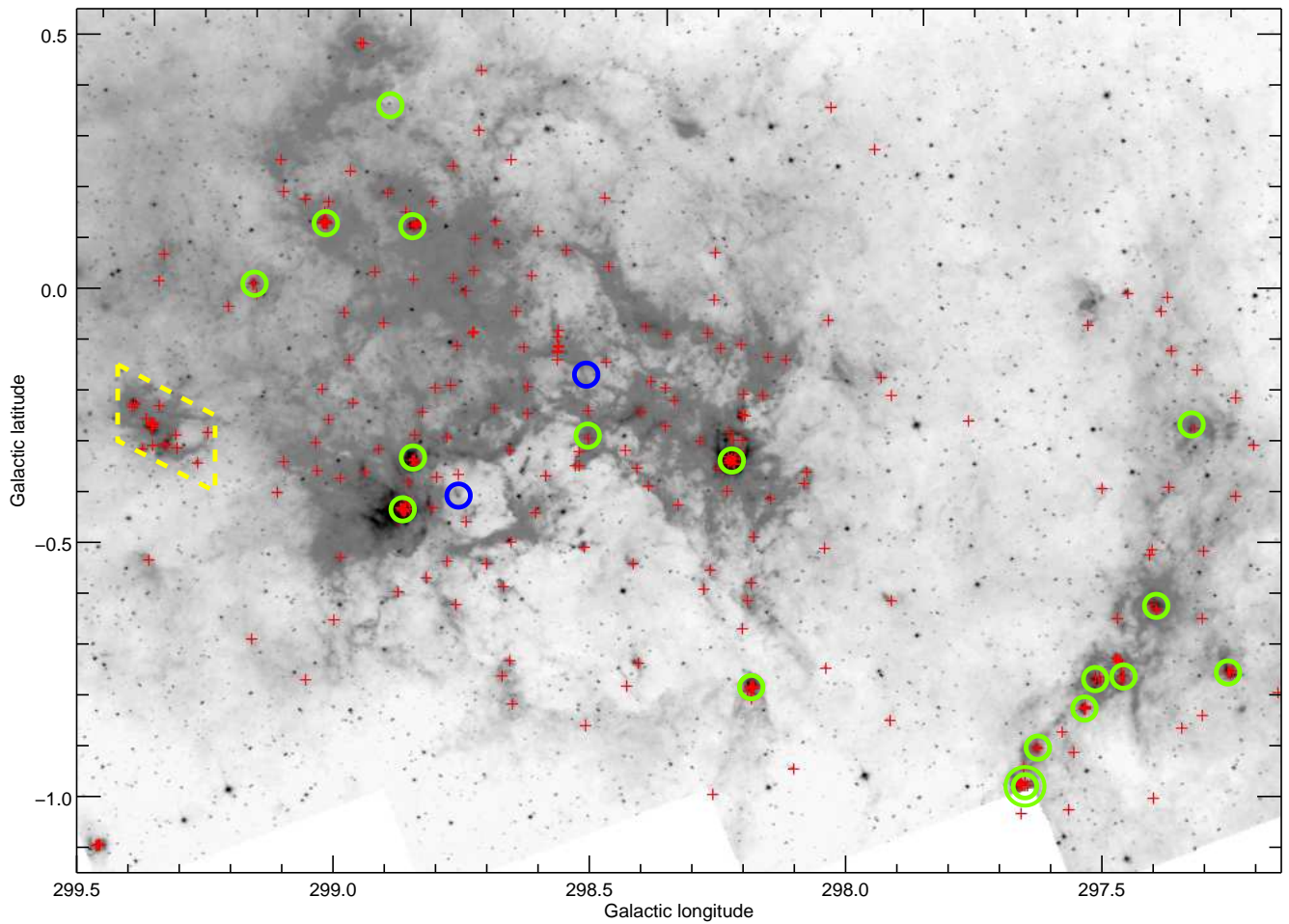}
        \caption{GLIMPSE image of the Dragonfish nebula in the $8 \mu \mathrm{m}$ band showing the locations of YSOs or candidate YSOs (red plus signs), as well as clusters or candidate clusters that are related to signs of star formation (green circles) or that host confirmed Wolf-Rayet stars (blue circles). The new cluster candidate we present in this paper is shown with a double green circle. The yellow dashed polygon encloses the foreground region RCW 64.
              }
      \label{fig:ysos_clusters}
    \end{figure*}

We  also carried out a literature search of young cluster candidates and confirmed young clusters in the Dragonfish star-forming complex. To avoid unrelated foreground clusters or spurious overdensities of stars, we  only selected objects with strong evidence of ongoing or recent star formation, as follows. First, we have looked for matches with objects listed in Table \ref{tab:dragonfish_vr}, or other signs of current star formation from the literature, e.g., extended green objects \citep[EGO;][]{cyganowski+08}.
Second, we  checked if strong $8 \mu \mathrm{m}$ emission is spatially coincident with the clusters in the GLIMPSE image, implying that the cluster is ionizing an \ion{H}{ii} region. Third, we  queried young stellar objects (YSOs) or candidate YSOs in the Dragonfish complex through the VizieR Catalogue Service (see Fig. \ref{fig:ysos_clusters}); we select candidate clusters that are spatially coincident with two or more YSOs or YSO candidates. And finally, we  checked if results of the already characterized clusters are consistent with membership to the Dragonfish star-forming complex in terms of ages and kinematics.

We remark the special case of  candidate clusters VVV CL012, La Serena 30, and La Serena 31 \citep{borissova+11, barba+15}. We  rejected these objects since their spatial distribution makes likely a physical association with the foreground \ion{H}{ii} region RCW 64.

Incidentally, we  found a compact clump of YSO candidates at $(l,b) \approx (297.65,-0.98)$, in Fig. \ref{fig:ysos_clusters}, surrounded by clouds and bubbles, and very close to the water maser \object{Caswell H2O 297.66-00.97} \citep{caswell+89}. This multiple evidence points to the existence of a previously undiscovered young embedded cluster. Therefore we report this object as a new cluster candidate in the Dragonfish complex.

\begin{table*}
  \caption{Confirmed and candidate clusters with strong evidence of membership of the Dragonfish star-forming complex}             
  \label{tab:clusters}      
  \centering          
  \begin{tabular}{l l l l l l}       
  \hline\hline                          
  ID        & $l$ ($\degr$)& $b$ ($\degr$)&  Status  &  Evidence                        &  Refs.  \\
  \hline
La Serena 17  &   297.254  &   $-$0.756  &  Candidate  &  \ion{H}{ii} region, YSO clustering, bubble          &          1 \\
La Serena 18  &   297.325  &   $-$0.268  &  Candidate  &  \ion{H}{ii} region                                    &          1 \\
Mercer 28   &    297.394 &  $-$0.625  &  Candidate & \ion{H}{ii} region, YSO clustering, maser                    &    7  \\
La Serena 19  &   297.458  &   $-$0.764  &  Candidate  &  \ion{H}{ii} region, YSO clustering                  &          1 \\
Mercer 29   &    297.513 &  $-$0.769  &  Candidate & \ion{H}{ii} region, YSO clustering                           &    7  \\
La Serena 20  &   297.534  &   $-$0.827  &  Candidate  &  \ion{H}{ii} region, YSO clustering                  &          1, 8 \\
La Serena 22  &   297.625  &   $-$0.904  &  Candidate  &  \ion{H}{ii} region, YSO clustering                  &          1 \\
New candidate &  297.65  &   $-$0.98    &   Candidate  &  \ion{H}{ii} region, YSO clustering, maser, bubbles   &      10 \\
DBSB 75     &    298.184 &  $-$0.785  &  Confirmed & \ion{H}{ii} reg., YSO clust., maser, $\varv_\mathrm{LSR} = 28.83$ km/s, age: 1 Myr &  4,5  \\
DBSB 74     &    298.222 &  $-$0.339  &  Candidate & \ion{H}{ii} region, YSO and maser clustering               &    4  \\
La Serena 24  &   298.503  &   $-$0.290  &  Candidate  &  \ion{H}{ii} region                                    &          1 \\
VVV CL011    &    298.506 &  $-$0.170  &  Confirmed & Massive stars, cluster parameters (see text)                &   2,3  \\
Mercer 30   &    298.755 &  $-$0.408  &  Confirmed & Massive stars, bubble, cluster parameters (see text)        &  6,7,10  \\
DBSB 129    &    298.844 &  $-$0.333  &  Confirmed & \ion{H}{ii} region, YSO clustering                         &   4,5  \\
La Serena 27  &   298.845  &   $+$0.122  &  Candidate  &  \ion{H}{ii} region, YSO clustering                  &          1 \\
Mercer 31   &    298.864 &  $-$0.435  &  Candidate & \ion{H}{ii} region, YSO clustering                          &    7  \\
La Serena 28  &   298.888  &   $+$0.360  &  Candidate  &  EGO clustering                                        &         1 \\
G3CC 2      &    299.014 &   $+$0.128    &  Candidate & YSO and maser clustering                                 &    8  \\
La Serena 29  &   299.153  &   $+$0.009  &  Candidate  &  \ion{H}{ii} region, YSO clustering                          &       1 \\
\hline                  
\end{tabular}
\tablebib{(1) \citet{barba+15}; (2) \citet{borissova+11}; (3) \citet{chene+13}; (4) \citet{dutra+03}; (5) \citet{kharchenko+13}; (6) \citet{kurtev+07}; (7) \citet{mercer+05}; (8) \citet{morales+13}; (9) \citet{solin+14}; (10) This work.
}
\end{table*}

After discarding clusters (or candidates) with no clear relation to the Dragonfish star-forming complex, we obtain 19 objects that are listed in Table \ref{tab:clusters}, including the new detection. As seen in Fig. \ref{fig:ysos_clusters}, thirteen of them show simultaneously two signs of ongoing clustered star formation: a crowded group of YSO detections and the presence of a dense cloud, as similarly observed in nearby star-forming regions \citep{gutermuth+11}. On the other hand, Table \ref{tab:clusters} lists two additional clusters that do not show any of the signs of current star formation (i.e., YSO, dense cloud, maser, or EGO), implying somewhat older ages. One of them is Mercer 30, which is undoubtedly associated with the Dragonfish complex in light of the thorough evidence we have presented in this paper, i.e. ages, radial velocities, spectrophotometric distances, and the presence of a bubble around the cluster. The other object is VVV CL011, whose membership is discussed below.
    
\citet{chene+13} carried out a spectroscopic study of VVV CL011, finding one WN9/OIf star and two or three early-B dwarfs (one of them is dubious owing to low S/N). The characterization yields an age between 3 and 7 Myr and a lower mass limit of $M_\mathrm{CL11} \geq (660 \pm 150)~\mathrm{M_\odot}$. These parameters, together with spectral types, lead us to interpret this cluster as a smaller sibling of Mercer 30. Unfortunately, proper motions or radial velocities could not be measured accurately, and the spectrophotometric distance is uncertain, given the low number of spectra and the wide range of resulting distances, which range from 4.65 to 10.29 kpc.
Nevertheless, the projected location of VVV CL011 at the central part of the Dragonfish Nebula, together with spectral types and the cluster age, hint at a real membership of this star-forming complex. In any case, the known hot massive stellar population and the low cluster mass (relative to Mercer 30) indicate that VVV CL011 is a minor ionization contributor that can be neglected when addressing the whole star-forming complex.

  \subsection{On the field massive stars}
  \label{sec:field}
  
A significant fraction of massive stars born in clusters are released to the field through cluster disruption \citep{lada-lada03,weidner+07}, which is enhanced by their location in massive star-forming complexes \citep{grosbol-dottori13}. Surviving clusters are also responsible for populating the surrounding field with runaway massive stars \citep{fujii-portegieszwart11,gvaramadze+12}. Thus, the presence of a wealth of clusters of young age ($\le 7$ Myr) on the Dragonfish complex should imply a rich population of field massive stars expelled by these processes.
  
    \begin{figure}
      \centering
        \includegraphics[width=\hsize, bb=20 10 480 340, clip]{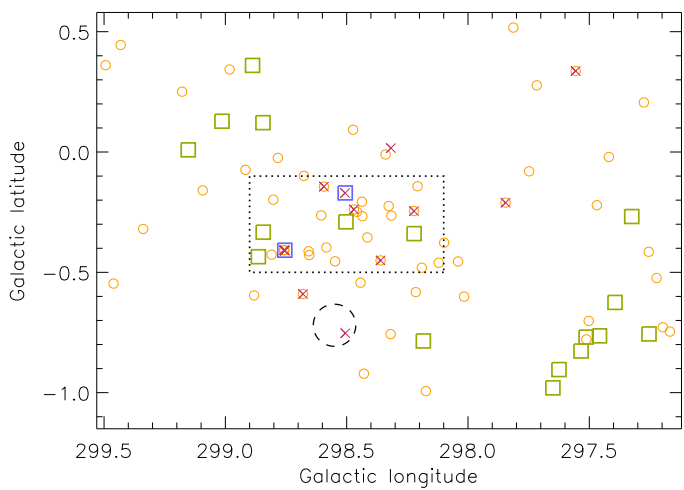}
        \caption{Location of the candidate WR stars we  found in the Dragonfish complex (orange circles) together with Wolf-Rayet stars from the literature (red crosses) and the same clusters that are highlighted in Fig. \ref{fig:ysos_clusters} (green/blue squares; the blue ones host confirmed WR stars). 32\% of clusters and 38\% of WR candidates, as well as the bulk of confirmed WR stars, are located within the dotted rectangle. The dashed circumference shows the position and size of the overdensity of 2MASS sources found by \citet{rahman+11a}.
              }
      \label{fig:wrclusters}
    \end{figure}

\citet{mauerhan+11} provided a simple but powerful method for locating field populations of massive stars. Although that paper was aimed at locating Wolf-Rayet stars, most  false positives are emission-line OB stars, which can also be considered as tracers of young populations. Four photometric (color-magnitude and color-color) diagrams were built by combining GLIMPSE and 2MASS magnitudes; the authors defined a certain region in each diagram that contains the candidate WR stars \citep[][see their Fig. 1]{mauerhan+11}.

We  applied the above  method to the whole region of the Dragonfish complex, yielding 58 objects that fulfill  all four selection criteria simultaneously. Their spatial distribution (Fig. \ref{fig:wrclusters}) is conspicuously concentrated in the central area $(l, b) \sim (298.5, -0.3)$, where the two clusters with confirmed Wolf-Rayet stars are located. In fact, 22 out of 58 WR candidates, as well as six out of 19 clusters (or candidate clusters), are located in the rectangle defined by $298.1 < l < 298.9; -0.5 < b < -0.1$, which approximately spans $1/10$ of the projected sky area towards the Dragonfish nebula. Strikingly, none of the massive star candidates is found close to the position where \citet{rahman+11b} allegedly confirmed the most luminous OB association in the Galaxy (see Fig. \ref{fig:wrclusters}).

To cross-check these WR candidates with already discovered WR stars, we  queried the SIMBAD astronomical database. We recover nine out of 58 WR candidates, however five additional objects that are catalogued as Wolf-Rayet stars were not found using the \citet{mauerhan+11} method. Since these five items include the object that were classified by \citet{rahman+11b} as a WN9 member of the Dragonfish Association, it is very important to investigate the causes of other non-detections. It turns out that three of them are located in very crowded regions of Mercer 30 and VVV CL011, where GLIMPSE has problems  resolving them, preventing the use of the color-selection method.
The remaining non-detection is \object{Hen 3-759}, which was included in the \citet{roberts62} catalog of Wolf-Rayet stars. However, \citet{crowther-evans09} reject this spectral type, reclassifying this object as O8 Iaf. Nevertheless, it is remarkable that Hen 3-759 fulfills two of the four \citet{mauerhan+11} color criteria and is very close to the region that defines WR candidates in the two remaining diagrams. In conclusion, star \#5 from \citet{rahman+11b} is the only catalogued Wolf-Rayet star whose absence in the results of our WR/Be candidate search remains unexplained so far.

After a similar WR candidate search in other star-forming regions, \citet{marston+13} carried out a spectroscopic follow-up of all the previously unconfirmed candidates, finding that 15\% of them were new WR stars. Assuming the same success rate on our sample of 49 new candidates (i.e. excluding the nine recovered WR stars), we expect $\sim 7$ undiscovered WR stars in the Dragonfish complex, which need to be added to the 7 already catalogued field WR stars. Owing to the very strong concentration of these young massive star candidates around the clusters with WR stars in the central part of the region (Fig. \ref{fig:wrclusters}), only a small minority of them are expected to be foreground or background stars.
Still, we will assume that the ratio of physically unrelated WR stars can be as high as 5/14. Thus, we reach the conservative approach that at least nine out of 14 expected WR stars in the sky area of the Dragonfish nebula are real members of the star-forming complex, i.e., three times the number of WR members of Mercer 30. If we also assume that the cluster and field populations of hot massive stars are homologous (i.e., with similar ratios of stellar types), we infer that the field population has an ionizing power at least three times that of Mercer 30: $Q^H_\mathrm{Field} \geq 2.01 \times 10^{51}~\mathrm{s^{-1}}$.

\begin{table*}
  \caption{Observed continuum fluxes at 5 GHz of \ion{H}{ii} regions, obtained from the literature, that are ionized by clusters (including the overall Dragonfish nebula), together with derived ionizing fluxes at 12.4 kpc.}             
  \label{tab:flux}      
  \centering          
  \begin{tabular}{l l l l l l }       
  \hline\hline 
\ion{H}{ii} region & Embedded population & $l$ ($\degr$)& $b$ ($\degr)$ & $f_{\nu}$ [Ref.] &  $Q^\mathrm{H}$ (12.4 kpc) \\
  \hline
GAL 297.51-0.77   & La Serena 19/20; Mercer 29 & 297.506 & $-$0.765   & 3.5 Jy [1]  & $8.6 \times 10^{49}~\mathrm{s^{-1}}$ \\
GAL 297.66-0.98   & La Serena 22; New candidate& 297.655 & $-$0.977   & 1.6 Jy [1]  & $3.9 \times 10^{49}~\mathrm{s^{-1}}$ \\
GAL 298.19-0.78   & DBSB 75             & 298.187 & $-$0.782   &        2.4 Jy [1]  & $5.9 \times 10^{49}~\mathrm{s^{-1}}$ \\
GAL 298.23-0.33   & DBSB 74             & 298.228 & $-$0.331   &        47.4 Jy [2]  & $1.159 \times 10^{51}~\mathrm{s^{-1}}$ \\
GAL 298.56-0.11   & Filament of YSOs    & 298.559 & $-$0.114   &        2.8 Jy [1]  & $6.8 \times 10^{49}~\mathrm{s^{-1}}$ \\
WMG70 298.8-0.3   & DBSB 129            & 298.838 & $-$0.347   &        16.0 Jy [4]  & $3.91 \times 10^{50}~\mathrm{s^{-1}}$ \\
GAL 298.87-0.43   & Mercer 31           & 298.868 & $-$0.423   &        42.4 Jy [2]  & $1.037 \times 10^{51}~\mathrm{s^{-1}}$ \\
GAL 299.02+0.15   & G3CC 2              & 299.016 & $+$0.148   &        2.6 Jy [1]  & $6.4 \times 10^{49}~\mathrm{s^{-1}}$ \\
  \hline
\multicolumn{2}{l}{Dragonfish Nebula} & 298.4 & $-$0.4  & $\lesssim 312$~Jy [3]\tablefootmark{a}    & $\lesssim 7.63 \times 10^{51}~\mathrm{s^{-1}}$ \\
  \hline                  
\end{tabular}
\tablefoot{
\tablefoottext{a}{The original estimate of 313 Jy has been slightly modified to account for the almost negligible foreground contribution (see text).}
}
\tablebib{(1) \citet{caswell-haynes87}; (2) \citet{conti-crowther04}; (3)\citet{murray-rahman10}; (4) \citet{wilson+70}.
}
\end{table*}

  \subsection{The ionization budget of the Dragonfish nebula}
  \label{sec:budget}
  
In sections \ref{sec:flux} and \ref{sec:field} we obtained the ionizing flux of Mercer 30 and a rough estimate for the contribution of field massive stars, respectively. Now, we will constrain the ionizing input from the remaining stellar population through the observed luminosities of \ion{H}{ii} regions that are powered by embedded clusters or associations. Specifically, the luminosity of an \ion{H}{ii} region within a star-forming complex is a lower limit of the contribution of the embedded massive stars to the overall nebular luminosity of the complex, since leakage of ionizing photons can affect a much wider envelope of low-density ISM \citep{anantharamaiah86, mckee-williams97, roshi-anantharamaiah01}, which produces the low-brightness extended emission that is observed across the Dragonfish complex at $8 \mu \mathrm{m}$.

The \ion{H}{ii} regions in the Dragonfish area with available measurements of H$^+$ free-free emission at the 5 GHz continuum are presented in Table \ref{tab:flux}, excluding the weak foreground object RCW 64, whose flux is 0.8 Jy \citep{caswell-haynes87}. In addition, Table \ref{tab:flux} shows the result for the overall Dragonfish nebula, which is an upper limit given that the foreground emitting material (which corresponds \textit{at least} to the RCW 64 contribution) would be subtracted from the 313 Jy estimate of \citet{murray-rahman10}.

All the \ion{H}{ii} regions listed in Table \ref{tab:flux} seem to host embedded clusters, except GAL 298.56-0.11, which still hosts a filamentary structure of YSOs that is clearly seen in Fig. \ref{fig:ysos_clusters}. To derive the luminosities at 12.4 kpc, we use the following simplified version of formulae in \citet{murray-rahman10} for the total number of Lyman-continuum photons per second:

\begin{equation}
  Q^\mathrm{H} \approx 1.59 \times 10^{47} d^2 f_\nu \,,
\end{equation}

where $d$ is the distance in kiloparsecs and $f_\nu$ is the measured flux in Janskys at 5 GHz. Adding all the \ion{H}{ii} regions, we obtain a lower limit of $Q^H_\mathrm{embedded} > 2.90 \times 10^{51} \mathrm{s}^{-1}$. The real value for the total ionizing luminosity emitted by embedded clusters may be significantly higher, given that this lower limit does not account for two contributions that we cannot assess. First, the aforementioned low-brightness emission in extended envelopes, which might absorb twice as many ionizing photons as  the dense \ion{H}{ii} regions, as argued by \citet{mckee-williams97}. Second, the Lyman-continuum photons emitted by other embedded cluster candidates with no available free-free emission measurements, namely Mercer 28 and La Serena 17/18/24/27/29.

\begin{table}
  \caption{Lower limits for the ionizing input provided by different stellar contributions, and fraction of the total photons needed to ionize the Dragonfish nebula}             
  \label{tab:budget}      
  \centering          
  \begin{tabular}{l l l}       
  \hline\hline 
 Stellar population   & $Q^H$ [$\mathrm{s^{-1}}$] & $Q^H/Q^H_{Dragonfish}$\\  
  \hline
 Mercer 30            & $6.70 \times 10^{50}$ & 0.09 \\
 VVV Cl011            & Vanished & $\sim 0$ \\
 11 embedded clusters\tablefootmark{a} & $2.90 \times 10^{51}$ & 0.38 \\
 6 embedded clusters  & Unknown & Unknown \\
 Field massive stars  & $2.01 \times 10^{51}$ & 0.26 \\
  \hline                  
\end{tabular}
\tablefoot{\tablefoottext{a}{Stellar populations presented in Table \ref{tab:flux}, including the filament of YSOs in GAL 298.56-0.11.}
}
\end{table}

Table \ref{tab:budget} shows a summary of the contributions to the ionizing input we have discussed in this paper. Adding all of them together, we find a lower limit of $Q^H_\star > 5.58 \times 10^{51} \mathrm{s}^{-1}$, or equivalently, $> 73\%$ of the photons needed to ionize the Dragonfish nebula (as calculated in Table \ref{tab:flux}, after subtracting the RCW 64 contribution to the overall free-free flux). We note that this is a lower limit, and the actual value must be well in excess of it, closer to 100\%. Therefore, our estimates make it difficult to explain the additional ionizing power of an alleged supermassive OB association in the southern part, as claimed by \citet{rahman+11a, rahman+11b}. If such an association exists, its contribution to the Dragonfish nebula ionization must be significantly lower than the sum of the two most luminous embedded clusters (DBSB 74 and Mercer 31; $\gtrsim 29\%$).

  \subsection{On the nature of the Dragonfish Association}
  \label{sec:naturedragon}

Since our data and discussion are in disagreement with the spectroscopic confirmation claimed by \citet{rahman+11b} for the so-called Dragonfish Association, we revisit  the evidence discussed by these authors. The low-resolution ($R \sim 1000$) H- and K-band observations presented in the above cited paper were designed to achieve a minimum S/N of 250, however inspection of the reduced spectra reveals significantly lower values. The final spectra of the two alleged luminous blue variables \citep[see Fig. 2 of][]{rahman+11b} have particularly poor S/N ($\sim 25$), causing ubiquitous absorption and emission peaks; some of these features, which are at the noise level, are labeled in that paper as \ion{Fe}{ii}, \ion{Fe}{i}, \ion{N}{i}, and \ion{Na}{i}.
On the other hand, the spectra presented by \citet{rahman+11b} show many absorption features that these authors interpret as interstellar lines of neutral metals, taking as a reference the ultraviolet observations of \citet{redfield-linsky04}. Although plenty of high-S/R infrared spectroscopic observations of early-type stars with similar or higher extinction values as the Dragonfish sources are available in the literature \citep[e.g.][]{figer+02, najarro+04, najarro+09, martins+07, martins+08, liermann+09, messineo+14, delafuente+15}, these lines have never been observed in H or K bands.

  \begin{figure}
      \centering
        \includegraphics[width=0.9\hsize]{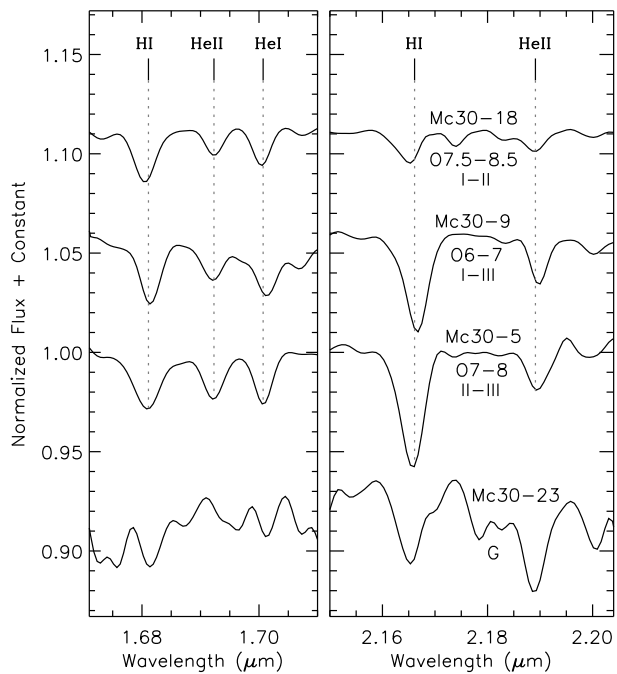}
        \caption{Mercer 30 O-type members, together with the late-type star Mc30-23, are shown in the same fashion as spectra presented by \citet{rahman+11b} for several objects in the Dragonfish Association. For an easier comparison, spectra have been artificially degraded to a resolving power of $\Delta \lambda / \lambda = 1000$.
              }
      \label{fig:degraded}
  \end{figure}

To simulate how O-type spectra in the Dragonfish complex would appear under the observing configuration of \citet{rahman+11b}, we  degraded the spectra of several O-type Mercer 30 members to a resolving power of $R = 1000$. In addition, we  performed the same task with the foreground G-type star Mc30-23. The resulting spectra are presented in Fig. \ref{fig:degraded} across the same wavelength ranges shown in \citet[Fig. 1]{rahman+11b}. Direct comparison between both figures makes evident that the alleged O-type spectra of the Dragonfish Association display very different features than O-type stars in Mercer 30, while showing a striking resemblance to the late-type features of Mc30-23.
Interestingly, the same authors had explained in their previous paper \citep{rahman+11a} that K-type giants with distances between 3.0 and 4.5 kpc would constitute photometric contaminants of the Dragonfish Association, since these objects could mimic the magnitudes and colors of more extinct and distant early-type massive stars. Therefore, the possibility that the stellar overdensity discovered by \citet{rahman+11a} is actually made up of these late-type giants deserves serious consideration and further research.

\section{Conclusions}

In this paper, we have presented a thorough analysis of hot luminous stars in Mercer 30, including a nearly complete sample of the post-main-sequence population. Our wealth of multi-epoch spectra and photometric data, together with \texttt{CMFGEN} quantitative modeling of cluster members, have yielded an accurate characterization of this YMC. Our new analysis has improved the accuracy of K-band extinction ($0.91 \pm 0.09$) and age ($(4.0 \pm 0.8)~\mathrm{Myr}$) compared with \citet{kurtev+07}. More importantly, other cluster parameters have undergone dramatic changes, illustrating the importance of analyzing the stellar population as extensively as possible. The main improvements on cluster characterization are summarized below.

The spectrophotometric distance has been increased from $(7.2 \pm 0.9)~\mathrm{kpc}$ to $(12.4 \pm 1.7)~\mathrm{kpc}$ based on OB stars and excluding WR types, as the latter lack a reliable K-band magnitude calibration. We have demonstrated that the new distance estimate is consistent with the radial velocity of Mercer 30, $(31 \pm 8)\mathrm{km~s}^{-1}$, using the Galactic rotation curve of \citet{brand-blitz93} (Fig \ref{fig:mc30_vgal}). This long distance has also prevented us from reaching any conclusive results based on proper motions.
The extensive photometry of VVV and the high spatial resolution of NICMOS/HST, along with the high number of spectroscopically observed stars, have enabled us to count stars above a certain luminosity limit, in such a way that the total cluster mass has been increased from the lower limit of $\sim 3000 M_\odot$ to $(1.6 \pm 0.6) \times 10^4 M_\odot$. These new results lead us to conclude that Mercer 30 is the most massive young cluster of the outer Milky Way ($R^\mathrm{(GC)} > 10 \mathrm{kpc}$) that has been discovered so far \citep[except if the stellar halo around h and $\chi$ Persei is added to the mass of the so-called Double Cluster;][]{currie+10}.

Our radial velocity and distance estimates mean we can place Mercer 30 in the Dragonfish  nebula which, in turn, is roughly located at the convex edge of the Saggitarius-Carina spiral arm. We have searched the whole Dragonfish complex for additional young massive stars, either in the field or as part of other young clusters of the Dragonfish complex, yielding 58 candidate or confirmed WR stars and 19 candidate or confirmed young clusters. The latter include Mercer 30 and a new candidate at $(l,b) \approx (297.65,-0.98)  $, which is reported here for the first time.
Although further analysis of this giant complex is required to accurately constrain  its properties, our evidence \citep[which is supplementary to][]{murray-rahman10} hints that this giant region might be one of the largest ($\sim$ 400 pc across) and most massive star-forming complexes in the Milky Way, being an analog of, for example, W43 \citep{nguyenluong+11}.

Membership of Mercer 30 to the Dragonfish star-forming complex has enabled us to use the cluster characterization results, specifically distance and Lyman-continuum flux ($Q^H_\mathrm{Mc30} \approx 6.70 \times 10^{50} \mathrm{s}^{-1}$), together with literature data, to probe the ionization budget for the whole Dragonfish nebula. The outcome leads us to conclude that the bulk of the ionizing input is provided by a variety of young massive stellar populations (namely Mercer 30, VVV CL011, several embedded clusters, and field stars), instead of the single supermassive OB association that was allegedly confirmed by \citet{rahman+11b}. As discussed in Section \ref{sec:naturedragon}, the stellar overdensity detected by \citet{rahman+11a} in the southern part of the Dragonfish complex might actually be a foreground feature that is composed of late-type stars; in any case, further observations are required to constrain its nature.

\begin{acknowledgements}

We wish to thank the anonymous referee for helping us improve the paper. We also thank John Hillier for providing the \texttt{CMFGEN} code. Part of this research has been supported by the Spanish Government through projects AYA2010-21697-C05-01, FIS2012-39162-C06-01, and ESP2013-47809-C3-1-R. D. dF. also acknowledges the FPI-MICINN predoctoral grant BES-2009-027786. We gratefully acknowledge use of data from the ESO Public Survey programme ID 179.B-2002 taken with the VISTA telescope, and data products from the Cambridge Astronomical Survey Unit. Support for JB, SRA and RK  is provided by the Ministry of Economy, Development, and Tourism's Millennium Science Initiative through grant IC120009, awarded to The Millennium Institute of Astrophysics, MAS. J. B. is supported by FONDECYT No.1120601, S. R. A. by FONDECYT No. 3140605.
This paper is partly based on observations made with the NASA/ESA \textit{Hubble} Space Telescope, obtained at the Space Telescope Science Institute, which is operated by the Association of Universities for Research in Astronomy, Inc., under NASA contract NAS 5-26555; these observations are associated with program \#11545. M.M.H. acknowledges support from ESO Chile Scientific Visitor Programme. This material is based upon work supported by the National Science Foundation under grant AST 10009550 to the University of Cincinnati. This work is based in part on data obtained as part of the UKIRT Infrared Deep Sky Survey. This research has made use of the SIMBAD database, operated at CDS, Strasbourg, France. This work is based in part on observations made with the \textit{Spitzer} Space Telescope, which is operated by the Jet Propulsion Laboratory, California Institute of Technology under a contract with NASA. This research has made use of the VizieR catalogue access tool, CDS, Strasbourg, France.
\end{acknowledgements}

\bibliographystyle{aa} 
\bibliography{paper2} 

\Online

\begin{appendix} 
\section{Details of \texttt{CMFGEN} models for Mercer 30.}
\label{sec:app}

\begin{figure*}
      \centering
      \includegraphics[angle=-90,width=0.98\hsize]{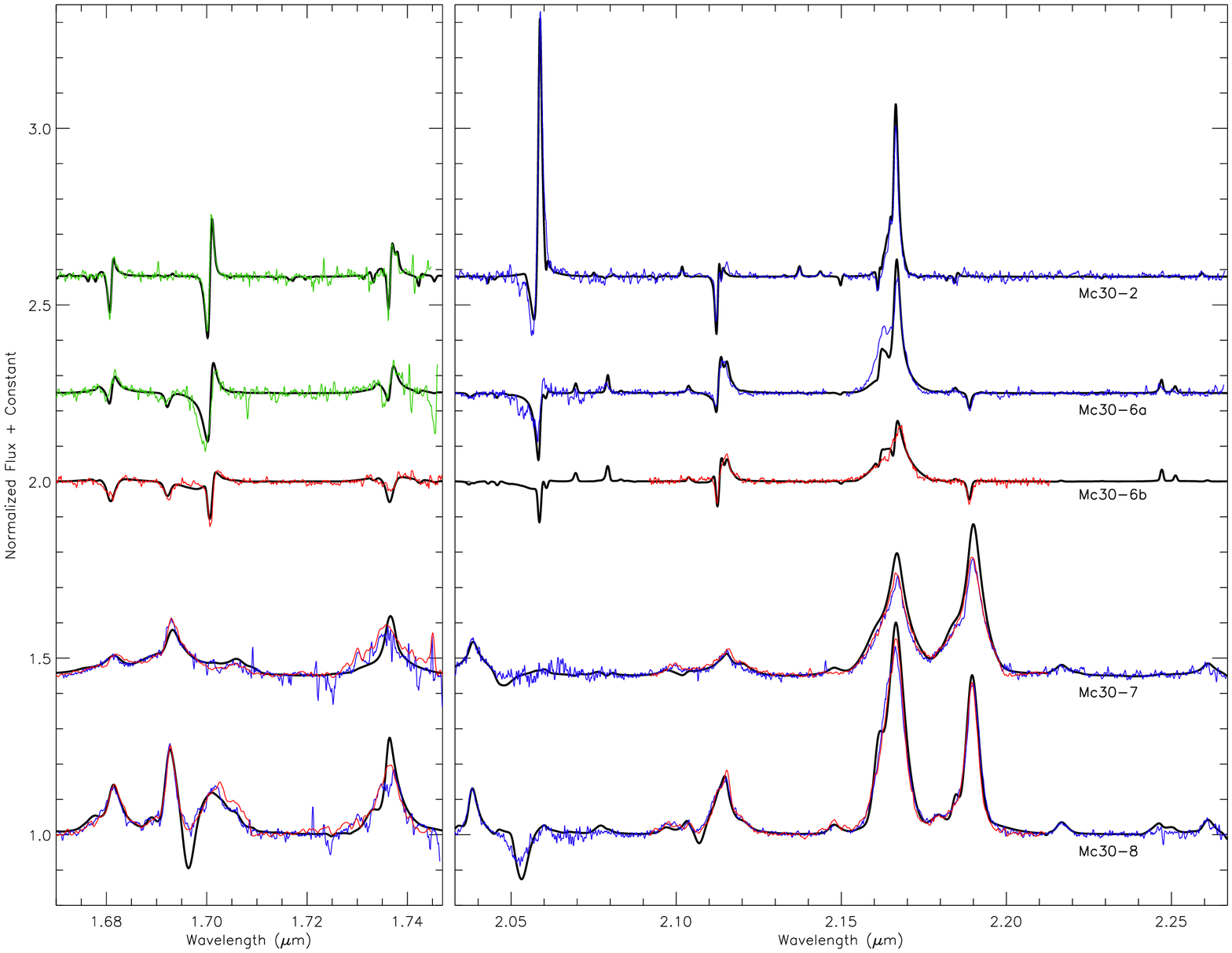}
      \caption{Models (black lines) fitted to the spectra that have been presented in the upper panels of Fig. \ref{fig:hotspectra}, following the same color code. Radial velocities of observed spectra have been corrected with the values of Tables \ref{tab:models} and \ref{tab:vr_variable}.
              }
    \label{fig:mod1}
  \end{figure*}

\begin{figure*}
      \centering
      \includegraphics[angle=-90,width=0.98\hsize]{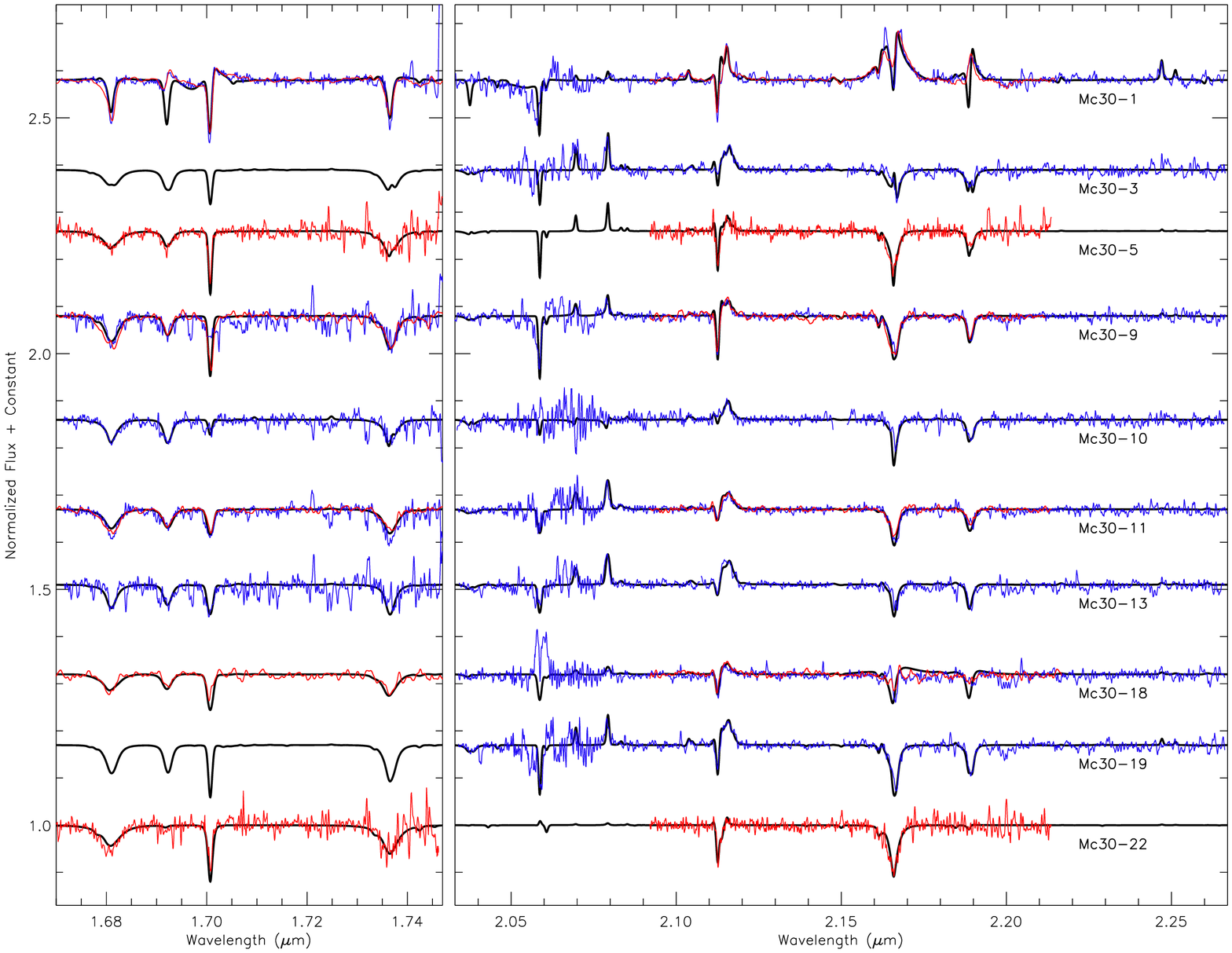}
      \caption{Models (black lines) fitted to the spectra that have been presented in the lower pnnels of Fig. \ref{fig:hotspectra}, following the same color code. Radial velocities of observed spectra have been corrected with the values of Tables \ref{tab:models} and \ref{tab:vr_variable}.
              }
    \label{fig:mod2}
  \end{figure*}

The full plots of model-fitting for cluster members are shown in Figs. \ref{fig:mod1} and \ref{fig:mod2}.
The important model parameters in Table \ref{tab:models} are complemented with those in Table \ref{tab:otherpars},
which are of secondary importance for our discussions. We display in
italics those  parameters that could not be observationally constrained, i.e.,
cases in which  $\varv\sin i$ and $\zeta_\mathrm{mac}$ 
are lower than the resolving power, or surface abundances of elements with no detected spectral features, 
or assumed terminal velocities for OB stars. The latter, however, are important for the $\dot M$ results,
since these are dependent on the $\varv_\infty$ values through the transformed radii \citep{schmutz+89}

\begin{table*}
  \caption{Parameters of \texttt{CMFGEN} models of Mercer 30 cluster members that are not crucial for our discussions.}             
  \label{tab:otherpars}      
  \centering          
  \begin{tabular}{l l l l l l l l l l l l l l l}       
  \hline\hline                          
ID&$\varv\sin i$\tablefootmark{a}&$\zeta_\mathrm{mac}$\tablefootmark{a}&$\xi_\mathrm{mic}$\tablefootmark{a}&$\varv_\infty$\tablefootmark{a}&$\beta$&$\dot M$~[$M_\odot$~yr$^{-1}$]&$CL_1$\tablefootmark{a}&$CL_2$\tablefootmark{a}\tablefootmark{b}&$CL_3$\tablefootmark{a}\tablefootmark{b}&$CL_4$\tablefootmark{b}& $Z_\mathrm{C}$\tablefootmark{c}&$Z_\mathrm{N}$\tablefootmark{c}&$Z_\mathrm{O}$\tablefootmark{c} \\
\hline
 Mc30-1 &\textit{65}&\textit{65}& 15 &         1400 & 1.35 & $2.2 \times 10^{-5}$ &   0.100  &  320  &   75   &   0.10    &$\leq$40  & 5.0 &$\leq$0.8 \\
 Mc30-2 &\textit{10}&    85     & 15 &         600  & 2.00 & $1.1 \times 10^{-5}$ &   0.250  &   10  &   60   &   0.15    &\textit{0.7}&\textit{11.4}&\textit{3.1}\\
 Mc30-3 &\textit{65}&\textit{65}& 15 & \textit{2500}& 1.05 & $7.3 \times 10^{-7}$ &   0.100  &   50  &   75   &   0.10    &   1.5    & 1.0  & 7.0 \\
 Mc30-5 &\textit{50}&\textit{50}& 12 & \textit{2500}& 0.80 & $1.2 \times 10^{-6}$ &\multicolumn{4}{c}{--- No clumping ---}&   1.0    & 0.8  & 2.5 \\
Mc30-6a & 35        &   110     & 20 &         1600 & 1.75 & $1.2 \times 10^{-5}$ &   0.100  &   65  &  100   &   0.20    &$\leq$0.6 & 4.0  &$\leq$3.0 \\
Mc30-6b &    100    &\textit{50}& 17 &         1600 & 0.90 & $9.4 \times 10^{-6}$ &   0.100  &  135  &   75   &   0.10    &$\leq$0.8 & 4.0  &$\leq$3.0 \\
 Mc30-7 &    180    &\textit{65}& 25 &         2000 & 1.00 & $2.0 \times 10^{-5}$ &   0.075  &  140  &    5   &   0.10    &$\leq$0.2 & 10.0 &$\leq$2.0 \\
 Mc30-8 &\textit{65}&\textit{65}& 25 &         1200 & 1.07 & $3.6 \times 10^{-5}$ &   0.250  &  135  &  170   &   1.00    &$\leq$0.2 & 19.5 &$\leq$0.6 \\
 Mc30-9 &\textit{65}&\textit{65}& 15 & \textit{2500}& 0.80 & $4.3 \times 10^{-6}$ &\multicolumn{4}{c}{--- No clumping ---}&   1.4    & 0.8  & 4.0 \\
Mc30-10 &\textit{65}&\textit{65}& 20 & \textit{2500}& 1.25 & $2.7 \times 10^{-6}$ &\multicolumn{4}{c}{--- No clumping ---}&   0.8    & 1.9  & 3.2\\
Mc30-11 &    150    &\textit{65}& 20 & \textit{2500}& 1.05 & $4.2 \times 10^{-6}$ &\multicolumn{4}{c}{--- No clumping ---}&   1.0    & 1.0  & 4.0 \\
Mc30-13 &    130    &\textit{65}& 25 & \textit{2500}& 1.50 & $2.7 \times 10^{-6}$ &\multicolumn{4}{c}{--- No clumping ---}&   0.4    & 1.5  & 5.2\\
Mc30-18 &    120    &    100    & 14 & \textit{2500}& 0.80 & $3.1 \times 10^{-6}$ &   0.300  &  150  &   75   &   0.30    &   0.4    & 1.2  & 1.5 \\
Mc30-19 &\textit{65}&\textit{65}& 17 & \textit{2500}& 1.05 & $2.0 \times 10^{-6}$ &\multicolumn{4}{c}{--- No clumping ---}&   0.9    & 3.5  & 4.6 \\
Mc30-22 &\textit{50}&    115    & 10 & \textit{2500}& 0.80 & $5.4 \times 10^{-7}$ &\multicolumn{4}{c}{--- No clumping ---}&\textit{1.5}&\textit{0.8}&\textit{4.0} \\
\hline
\end{tabular}
\tablefoot{\tablefoottext{a}{The $\varv\sin i$, $\zeta_\mathrm{mac}$, $\xi_\mathrm{mic}$, $\varv_\infty$, $CL_2$, and $CL_3$ values are in $\mathrm{km~s}^{-1}$.} \tablefoottext{b}{The coefficients for the clumping law of \citet{najarro+09} are used, however when $CL_4=CL_1$ the clumping law of \citet{hillier-miller99} is recovered with $a=CL_1$ and $b=CL_2$ (which occurs for the Mc30-1, Mc30-3, Mc30-6b, and Mc30-18 models).} \tablefoottext{c}{Metal abundances are expressed in mass fraction and in units of $10^{-3}$.}
}
\end{table*}

\subsection{Notes for individual objects}

\subsubsection{Mc30-1}
\label{asec-mc30-1}
Mc30-1 shows clear evidence of binarity through both radial velocity and line-profile variations. The composite spectrum indicates the presence of a late-type O supergiant, which cause the strong \ion{He}{i} lines and a mid-Of, the latter displaying significant He enrichment and producing the \ion{He}{ii} emission features. As discussed in section \ref{sec:age} the large luminosity of Mc30-1 is also consistent with a binary nature. Since further observations are required for a proper characterization of the binary system and subsequent disentangling of the spectra, we opted for a compromise model (Fig.\ref{fig:mod2}) which fits the combined spectra. We note that if the \ion{He}{ii} lines are ignored, we can obtain an almost perfect fit of the \ion{H}{i}, \ion{He}{i}, and \ion{N}{iii}/\ion{C}{iii} lines in the H- and K-Bands with a 2000~K cooler model with no He enrichment.

\subsubsection{Mc30-2}
\label{asec-mc30-2}
Mc30-2 stellar properties indicate a hypergiant nature for this object. Although He is only mildly enhanced ($\mathrm{He}/\mathrm{H}=0.15$, the star has a very dense wind, displaying P-Cygni type profiles even for the high Brackett-series members in the H band. We obtain moderate nitrogen enhancement and carbon depletion from the \ion{N}{ii} and \ion{C}{ii} lines, respectively. Regarding the $\alpha$-elements, we obtain
solar Si abundance , while a twice solar abundance is derived for Mg.

\subsubsection{Mc30-3}
\label{asec-mc30-3}
Our best model of the K-Band spectrum of this O6If star is consistent with very mild chemical evolution.
We find no He enrichment, small depletion for C, a slight enrichment for N, and no depletion for O 
(which vastly dominates the $2.115 \mu \mathrm{m}$ feature). Our results are consistent with an initial solar
oxygen abundance. The complex shapes of the Brackett-$\gamma$ and \ion{He}{ii}~$2.189 \mu \mathrm{m}$ lines are
satisfactorily reproduced and are consistent with a clumping factor of
$CL_1 = 0.1$.

\subsubsection{Mc30-5}
\label{asec-mc30-5}
We find a relatively strong wind for the luminosity
class IV-V of this object.

\subsubsection{Mc30-6}
\label{asec-mc30-6}

\paragraph{Mc30-6a.} The observations in the H and K bands where Mc30-6a
dominates the spectrum were taken at different epochs and slit
orientations. This is reflected in the model fits, which attempt to
match the strongest features (a higher weight has been placed for the
K-Band lines) but clearly fail to reproduce some of the blue absortion features in
the H-Band profiles. 

\paragraph{Mc30-6b.} A much better fit is obtained for this simultaneous observational
setting, dominated by Mc30-6b, taken with the same slit orientation and therefore same
contribution of both objects.
Both models indicate the presence of CNO-processed material at the
surface of these objects.

\subsubsection{Mc30-7}
\label{asec-mc30-7}
Mc30-7 displays large variations 
in $\varv_\mathrm{LSR}$. Apart from the weak feature at $2.095 \mu
\mathrm{m}$, which could be due to \ion{N}{v} and hence, indicate the
presence of a much hotter object, no other traces of a putative
companion are found.  While
preserving the observed  
Br-$\gamma$ vs  \ion{He}{ii}~$2.189 \mu \mathrm{m}$ ratio, 
our best, compromise model produces
slightly stronger line profiles. A lower wind density would
significantly underestimate the rest of H- and K-band hydrogen and helium lines. 
The derived He and CNO abundaces are consistent with a WNL (WNh) evolutionary
phase for this object.

\subsubsection{Mc30-8}
\label{asec-mc30-8}
Our model reproduces quite satisfactorily the observed lines, except
for the absorption component of the \ion{He}{i}~$1.700 \mu \mathrm{m}$
line. Like Mc30-7, we obtain He and CNO patterns, corresponding to a
WNL phase. 

\subsubsection{Mc30-9}
\label{asec-mc30-9}
No traces of binarity are found in the observed line profiles, though
 clear variations are detected in 
 $\varv_\mathrm{LSR}$ for Mc30-9.
 The observed \ion{He}{i}~$1.700 \mu \mathrm{m}$ line is clearly affected
 by poor cancellation of the strong OH sky line. 
 The H- and K-band spectra are very well reproduced by our best,
 unclumped model. 
 Our derived CNO pattern is consistent with the rest of O I-III stars
 in the sample.

\subsubsection{Mc30-10}
\label{asec-mc30-10}
Despite the variable  $\varv_\mathrm{LSR}$ displayed by the object, we
obtain a very good fit to the mean spectrum. 
In the parameter domain where Mc30-10 is located,
as the temperature is increased, the
\ion{C}{iv}~$2.07/8 \mu \mathrm{m}$ 
lines move from emission to absorption. Therefore, our models are
consistent with the non-detection of these lines.
The \ion{CNO}{iii}~$2.115 \mu \mathrm{m}$ feature is basically
dominated by \ion{O}{iii}, implying a $\sim 0.7 \times~$ solar oxygen
abundance.

\subsubsection{Mc30-11}
\label{asec-mc30-11}
We obtain an excellent fit to the H- and K-Band spectra with an
unclumped model. We find mild C depletion and N enrichment and obtain
a $\sim 0.8 \times$solar oxygen abundance.

\subsubsection{Mc30-13}
\label{asec-mc30-13}
This object displays variable  $\varv_\mathrm{LSR}$ and traces of
composite nature in the He lines of the H-Band. Nevertheless, a very
good general fit is obtained to the H- and K-Band spectra of Mc30-13.
The shapes of the Brackett lines are well reproduced with a moderate
$\beta$ and with clumping starting relatively close to the photosphere. 
We derive mild CN processing.

\subsubsection{Mc30-18}
\label{asec-mc30-18}
This object has also variable  $\varv_\mathrm{LSR}$ and the 
Br-$\gamma$ and  \ion{He}{ii}~$2.189 \mu \mathrm{m}$ profile shapes
display a composite nature which is confirmed by our model fits. A
lower temperature companion would provide the required dilution to
explain our derived low CNO
values as well as the observed strong \ion{He}{i} components in the
Brackett lines.

\subsubsection{Mc30-19}
\label{asec-mc30-19}
We obtain an excellent fit to the K-Band observations with an
unclumped model. 
$T_\mathrm{eff}$ is fairly well constrained by the 
\ion{He}{i}~$2.112 \mu \mathrm{m}$ absorption and 
\ion{He}{ii} lines at $2.189$ and $2.037\mu \mathrm{m}$. The strength of the 
\ion{CNO}{iii}~$2.115 \mu \mathrm{m}$ feature, together with 
\ion{C}{iv}~$2.07/8 \mu \mathrm{m}$ and \ion{O}{iii}~$2.104 \mu
\mathrm{m}$ are consistent with moderate CN processing and $\sim 0.8$
solar oxygen abundance. 

\subsubsection{Mc30-22}
\label{asec-mc30-22}
The \ion{He}{ii}~$2.189 \mu \mathrm{m}$ line is not detected and a
very weak feature is observed at the position of \ion{He}{ii}~$1.693
\mu \mathrm{m}$ line. The latter, together with the presence of the \ion{He}{i}~$2.113 \mu
\mathrm{m}$ in absorption constrain the stellar temperature. We find that the uncertainty in $T_\mathrm{eff}$ towards lower values 
is larger ($-2500 \mathrm{K}$) than discussed in Section \ref{sec:modeling}. The $\log~g$ uncertainty is also relatively high
($\pm 0.2\mathrm{dex}$). Clumping is not required to match the observations.
No metal lines are detected, so only upper limits can be
placed on CNO abundances (see Table~\ref{tab:otherpars}).

\end{appendix}

\end{document}